\documentclass[a4paper,11pt]{article}
\pdfoutput=1
\usepackage{geometry}
\usepackage{a4wide,slashbox}
\usepackage{graphicx}
\usepackage{epsf}
\usepackage{amsmath}
\usepackage{amssymb}

\usepackage{cite}
\usepackage{multirow,tabularx}
\usepackage{appendix}
\usepackage{tikz}
\usepackage{graphicx,amsmath,amsfonts,amssymb,amsthm,euscript,braket,xcolor}
\newcommand{\be}{\begin{equation}}
\newcommand{\ee}{\end{equation}}

\newcommand{\Rmnum}[1]{\expandafter\@slowromancap\romannumeral #1@}
\newcommand{\bea}{\begin{eqnarray}}
\newcommand{\eea}{\end{eqnarray}}

\newcommand{\p}{\partial}
\numberwithin{equation}{section}

\newcommand*\circled[1]{\tikz[baseline=(char.base)]{
            \node[shape=circle,draw,inner sep=2pt] (char) {#1};}}

\begin{document}

\title{\bf Interplay between the holographic QCD phase diagram and entanglement entropy}
\author{\textbf{David Dudal$^{a,b}$}\thanks{david.dudal@kuleuven.be},
\textbf{Subhash Mahapatra$^{c}$}\thanks{mahapatrasub@nitrkl.ac.in},
 \\\\\
\textit{{\small $^a$ KU Leuven Campus Kortrijk -- Kulak, Department of Physics, Etienne Sabbelaan 53 bus 7657,}}\\
\textit{{\small 8500 Kortrijk, Belgium}}\\
\textit{{\small $^b$  Ghent University, Department of Physics and Astronomy, Krijgslaan 281-S9, 9000 Gent, Belgium}}\\
\textit{{\small $^c$ Department of Physics and Astronomy, National Institute of Technology Rourkela, Rourkela - 769008, India}}}
\date{}

\maketitle
\abstract{In earlier work, we introduced a dynamical Einstein--Maxwell--dilaton model which mimics essential features of QCD (thermodynamics) below and above deconfinement. Although there are some subtle differences in the confining regime of our model as compared to the standard results, we do have a temperature dependent dual metric below $T_c$ as well, allowing for a richer and more realistic holographic modeling of the QCD phase structure. We now discuss how these features leave their imprints on the associated entanglement entropy when a strip region is introduced in the various phases. We uncover an even so rich structure in the entanglement entropy, consistent with the thermodynamical transitions, while again uncloaking some subtleties. Thanks to the temperature dependent confining geometry, we can present an original quantitative prediction for the phase diagram in terms of temperature and strip length, reporting a critical end point at the deconfinement temperature. We also generalize to the case with chemical potential.}

\section{Introduction}
Two of the most promising avenues where the powerful ideas of gauge/gravity duality \cite{Maldacena:1997re,Gubser:1998bc,Witten:1998qj} can be used to obtain important physical results are entanglement entropy and quantum chromodynamics (QCD). In this paper, following up on the seminal work that merged these two avenues, \cite{Klebanov0709}, we will further investigate how the concept of entanglement entropy endows the QCD phase diagram with an extra relevant external parameter, roughly corresponding to the size of the entangling subsystem.  \\

The entanglement entropy, a measure of entanglement which simply means how different subsystems of a full quantum system are correlated, has been the subject of intense investigations in the high energy, condensed matter, quantum information etc.~community in recent years, and forms the basis for many applications in these diverse areas of physics. For example, in condensed matter systems, it has been suggested as an order parameter to characterize different quantum phases  \cite{Vidal:2002rm,Osborne:2002zz} whereas in gravity it has been suggested as a natural candidate to explain the Bekenstein--Hawking black hole entropy \cite{Bombelli:1986rw,Srednicki:1993im}. From a holographic point of view, a remarkably successful conjecture regarding the entanglement entropy was suggested in the seminal work of Ryu--Takayanagi \cite{Ryu:2006bv,Ryu:2006ef}, with recent proofs of the conjecture appearing in \cite{Casini:2011kv,Lewkowycz:2013nqa}. This conjecture geometrizes the concept of entanglement entropy and provides a useful bridge connecting gravity with many body quantum systems. The Ryu--Takayanagi holographic conjecture has stood the test of time by reproducing independent known results of the entanglement entropy and there are also strong indications that this conjecture might be relevant to understand quantum gravity as well \cite{VanRaamsdonk:2010pw,Balasubramanian:2013lsa}.\\

On the other hand, QCD is the well-tested theory of sub-atomic particles carrying strongly interacting color charge, which at low temperatures and densities are bound together in colorless hadronic bound states due to confinement. Understanding QCD properties and its phase structure are of upmost importance in high energy physics. The failure of conventional perturbative techniques and the sparse availability of genuinely non-perturbative methods however have hindered our understanding of QCD at strong coupling. Here, the idea of gauge/gravity duality again provides an efficient and elegant method by which not only the strongly coupled regime of QCD can be explored but also its dynamics can be studied, even in circumstances not readily accessible to lattice QCD simulations, still the most reliable tool to probe QCD physics in a controllable non-perturbative setting.\\

One of the main and original objectives of the gauge/gravity duality is to understand strongly coupled gauge theories such as QCD \cite{Witten9803}. In particular, constructing a gravity theory capable of describing properties of real QCD and from which testable predictions can be extracted is of importance, both to support or complement other takes on the same problem, coming from e.g.~lattice QCD, Dyson-Schwinger or Functional Renormalization Group equations, effective QCD models, etc.  By now many holographic QCD models, both top-down from string theory as well as phenomenological bottom-up models, which can correctly reproduce a large number of QCD properties, have been constructed, see \cite{Witten9803,Polchinski0003,Sakai0412,Sakai0507,Kruczenski0311,Karch0205,Klebanov0007,Erlich,Gursoy,Gursoy:2010fj,Jarvinen:2015ofa,Herzog0608,Karch0602,Karch1012,Colangelo:2011sr,Callebaut:2011ab,Callebaut:2013ria,Dudal:2015wfn,Dudal:2014jfa,Dudal:2018rki,
Braga:2017bml,Braga:2017fsb,Li:2016gfn,Fang:2015ytf,Andreev:2006nw,Andreev:2007rx,
Critelli:2016cvq,Giataganas:2017koz,Gherghetta:2009ac,Panero:2009tv,Cai:2012xh,Noronha:2010hb,Paula,He:2013qq,Yang:2015aia,Arefeva:2018hyo,Knaute:2017opk} for a necessarily incomplete selection. \\

Due to severe computational difficulties, present both at analytical as well as at numerical level, the discussion of entanglement entropy in QCD-like theories has been rather limited.  Most investigations are based on the holographic conjecture of Ryu--Takayanagi, as otherwise it is near to impossible to get a reliable non-perturbative estimate of the entanglement entropy relevant for QCD, with the exception of a few lattice oriented papers, see \cite{Buividovich:2008kq,Buividovich:2008gq,Itou:2015cyu}. In \cite{Klebanov0709}, it was first observed that the entanglement entropy can act as a probe of confinement in gravity duals of large $N$ gauge theories. In particular, a phase transition between connected and disconnected entangling surfaces as a function of size of the entangling surface, at which the order of the entanglement entropy changes, was suggested as a signature of (de)confinement. This idea was then generalized to a variety of confining systems \cite{Fujita0806,Kola1403,Kim,Lewkowycz,Ghodrati,Knaute:2017lll,Ali-Akbari:2017vtb}, and has moreover received numerical confirmation from the aforementioned lattice papers \cite{Buividovich:2008gq,Buividovich:2008kq,Itou:2015cyu}, see also \cite{Anber:2018ohz}. In \cite{Dudal:2016joz}, we performed a similar analysis including a background magnetic field and showed for the first time an anisotropic footprint of confinement/deconfinement transition in the entanglement entropy.\\

Most studies of the entanglement entropy in holographic theories have been restricted either to top-down or to phenomenological soft wall models of QCD. However, as it is well known, both of these models can face several limitations in describing real QCD. For example, the boundary theory of top-down gauge/gravity models generally contains conformal symmetries as well as additional sectors in its Hilbert space, coming from the Kaluza--Klein modes of extra dimensions \cite{Sakai0412,Sakai0507}, whose counterparts in QCD do not exist. On the other hand soft wall models, apart from being sometimes explicitly inconsistent with the Einstein equations, do not always exhibit the area law of the Wilson loop expectation value \cite{Karch1012} or experience issues to produce a meaningful chiral condensate \cite{Colangelo:2011sr,Dudal:2015wfn}. Most of these difficulties, however, can be overcome by more phenomenological ``bottom-up'' holographic QCD models where one constrains the five dimensional gravity theory in an ad-hoc way, based on physical intuition, as to reproduce the desirable properties of the four-dimensional boundary theory resembling QCD. It is thus of interest to investigate the entanglement entropy in such self-consistent bottom-up holographic QCD models, in particular to unravel further the extra insights it can offer into the QCD vacuum structure and confinement mechanism.\\

The main objective of this work is thus to fill the above mentioned gap by investigating the entanglement entropy in a recently developed holographic QCD model. For this purpose, we consider the Einstein--Maxwell--dilaton (EMD) holographic setup of \cite{Dudal:2017max} \footnote{Various other EMD like holographic models were considered earlier in the literature, for example see \cite{Gubser:2008yx,Gubser:2008ny,DeWolfe:2011ts,Rougemont:2015ona,Cai:2012xh,Noronha:2010hb,Paula,He:2013qq,Yang:2015aia,
Knaute:2017opk}.}. The novelty of the model lies in the fact that, apart from being analytically solvable, one can obtain a rich holographic QCD phase structure by tuning the scale factor $A(z)$ and/or dilaton potential (see eq.~(\ref{metsolution1})) while at the same time mimicking essential features of (lattice) QCD. Moreover, this model allows us to study temperature dependent properties of various observables not only in the deconfined but in the confined phase as well. In particular, in \cite{Dudal:2017max} we showed that by taking various forms of $A(z)$ one can obtain the standard confined and deconfined phases, as well as the novel and somewhat strange ``\textit{specious-confined}'' phase.  The latter phase, which is dual to a small black hole phase on the gravity side and hence explicitly includes the notion of temperature, corresponds to a boundary phase which does not strictly coincide with the standard confined phase (see \cite{Dudal:2017max} or later in this paper), nevertheless it shares many of its properties. For instance, the free energy and entropy of a probe quark-antiquark pair in this specious-confined phase are in qualitative agreement with ruling lattice QCD estimates. Moreover, an observation that is hard to come by in most holographic models which usually lack the notion of a temperature dependent confined phase: the specious-confined phase predicts a finite string tension at the deconfinement transition temperature, a finding supported by lattice QCD as well \cite{Cardoso:2011hh}.  Because of these remarkable properties of the dual boundary theory, one might wonder how this specious confined phase will reflect upon the entanglement entropy.\\

In this paper, following \cite{Dudal:2017max}, we again consider two different forms for $A(z)$. The first form $A_{1}(z)$ (see eq.~(\ref{Aansatz1})) gives a thermal-AdS/black hole phase transition on the gravity side, which on the dual boundary side corresponds to standard confinement/deconfinement phase transition. We then discuss the holographic entanglement entropy in these confined/deconfined phases  using a strip geometry of length $\ell$ as subsystem \cite{Ryu:2006bv,Ryu:2006ef}. Interestingly, the entanglement entropy in the confined phase exhibits the same features as suggested by \cite{Klebanov0709}. In particular, we again find a connected to disconnected entangling surface transition at some critical strip length $\ell_{c}$, at which the order of the entanglement entropy changes. However, in the deconfined phase no such transition exists as the entanglement entropy of the connected surface is always smaller than the disconnected surface. We then establish the QCD phase diagram by studying the entanglement entropy in the temperature-chemical potential plane. On the other hand, the second and more interesting form $A_{2}(z)$ (see eq.~(\ref{Aansatz2})) instead provides for a small/large black hole phase transition, which on the dual boundary side corresponds to the specious-confined/deconfined phase transition. In this case, the entanglement entropy of the deconfined phase is similar to the deconfined phase entanglement entropy obtained using $A_{1}(z)$. We have checked for several other forms of $A(z)$ as well  and found similar results for the dual deconfined phase.  However, the entanglement entropy in the specious-confined phase exhibits many new and interesting features. In particular, there is no connected to disconnected surface transition, however, now a novel connected to connected surface transition appears that we study in detail, making use of the entropic $\mathcal{C}$-function. Interestingly, for larger subsystems, the variation of entanglement entropy with respect to the strip length is extremely small (this variation is zero in the standard confined phase). These results further highlights the differences as well as similarities between standard confined and specious-confined phases, see also \cite{Dudal:2017max}. As such, we can draw the QCD phase diagram in terms of temperature and strip length, at least in the confined regime. Our findings here not only provide a strong support for the existence of a \textit{conjectured} line of phase transitions in the $(T,\ell_{c})$ plane, as first suggested in \cite{Buividovich:2008kq}, but lends a support for the existence of a similar line of phase transitions in the presence of a chemical potential as well, albeit with larger $\ell_c$ value. Further, we examine the specious-confined/deconfined QCD phase diagram by studying the entanglement entropy in the temperature-chemical potential plane.
\\

This paper is organised as follow. In the next section, we briefly survey our Einstein--Maxwell--dilaton holographic model. In section 3, we derive the necessary formulae for the entanglement entropy computation.  In section 4, using the first form of $A(z)$, we first examine the thermodynamics of the gravity solution and then discuss the entanglement entropy in the corresponding confined and deconfined phases. In section 5, we repeat the calculations of section 3 with the second form of $A(z)$ and study the entanglement entropy in the specious-confined and deconfined phases. Finally, we end this paper with conclusions and an outlook to future research
in section 5.

\section{Einstein--Maxwell--dilaton gravity vs.~boundary QCD}
The EMD holographic model at zero and finite temperature has been discussed thoroughly in \cite{Dudal:2017max}, including its analytic solution, and we refer the reader to \cite{Dudal:2017max} for more details. In this section, we briefly describe this model and present only the relevant analytic expressions, which will be important for our discussion in later sections.\\

The EMD holographic action in five dimensions consists of a field strength tensor $F_{MN}$ and dilaton field $\phi$ on the top of usual Einstein--Hilbert term,
\begin{eqnarray}
&&S_{EM} =  -\frac{1}{16 \pi G_5} \int \mathrm{d^5}x \sqrt{-g} \ \ \left[R-\frac{f(\phi)}{4}F_{MN}F^{MN} -\frac{1}{2}\partial_{M}\phi \partial^{M}\phi -V(\phi)\right],
\label{actionEF}
\end{eqnarray}
where $f(\phi)$ is a gauge kinetic function which represents the coupling between the gauge field $A_{M}$ and $\phi$. $V(\phi)$ is the potential of the dilaton field, whose explicit form is not required, and $G_5$ is the Newton constant in five dimensions. Interestingly, see also \cite{Yang:2015aia}, using the following Ans\"atze for the metric, $A_{M}$ and $\phi$,
\begin{eqnarray}
& & ds^2=\frac{L^2 e^{2 A(z)}}{z^2}\biggl(-g(z)dt^2 + \frac{dz^2}{g(z)} + dy_{1}^2+dy_{3}^2+dy_{3}^2 \biggr)\,, \nonumber \\
& & A_{M}=A_{t}(z), \ \ \ \ \phi=\phi(z)\,,
\label{metric}
\end{eqnarray}
the equations of motion of the EMD model can be solved analytically in terms of a single scale function $A(z)$,
{
\allowdisplaybreaks
\begin{eqnarray}
&&g(z)=1-\frac{1}{\int_{0}^{z_h} dx \ x^3 e^{-3A(x)}} \biggl[\int_{0}^{z} dx \ x^3 e^{-3A(x)} + \frac{2 c \mu^2}{(1-e^{-c z_{h}^2})^2} \det \mathcal{G}  \biggr],\nonumber \\
&&\phi'(z)=\sqrt{6(A'^2-A''-2 A'/z)}, \nonumber \\
&& A_{t}(z)=\mu \frac{e^{-c z^2}-e^{-c z_{h}^2}}{1-e^{-c z_{h}^2}}, \nonumber \\
&& f(z)=e^{c z^2 -A(z)}\,, \nonumber \\
&&V(z)=-3L^2z^2ge^{-2A}\left[A''+A' \bigl(3A'-\frac{6}{z}+\frac{3g'}{2g}\bigr)-\frac{1}{z}\bigl(-\frac{4}{z}+\frac{3g'}{2g}\bigr)+\frac{g''}{6g} \right]\,,
\label{metsolution1}
\end{eqnarray}}
where
\[
\det \mathcal{G} =
\begin{vmatrix}
\int_{0}^{z_h} dx \ x^3 e^{-3A(x)} & \int_{0}^{z_h} dx \ x^3 e^{-3A(x)- c x^2} \\
\int_{z_h}^{z} dx \ x^3 e^{-3A(x)} & \int_{z_h}^{z} dx \ x^3 e^{-3A(x)- c x^2}
\end{vmatrix}.
\]
The (Einstein frame) gravity solution in eq.~(\ref{metsolution1}) corresponds to a black hole with horizon at $z=z_h$. To derive this solution, we have used the boundary condition that $g(z)$ goes to $1$ at the asymptotic boundary $z=0$, and that at the horizon $g(z_h)=0$. Here $\mu$ is the chemical potential of the boundary theory, which is obtained from the asymptotic boundary expansion of the gauge field. The Bekenstein--Hawking entropy and Hawking temperature of the black hole solution are given by,
\begin{eqnarray}
&&\hspace{-5mm}T= \frac{z_{h}^3 e^{-3 A(z_h)}}{4 \pi \int_{0}^{z_h} dx \ x^3 e^{-3A(x)}} \biggl[ 1+\frac{2 c \mu^2 \bigl(e^{-c z_h^{2}}\int_{0}^{z_h} dx \ x^3 e^{-3A(x)}-\int_{0}^{z_h} dx \ x^3 e^{-3A(x)}e^{-c x^{2}} \bigr)}{(1-e^{-c z_h^{2}})^2} \biggr]\,,  \nonumber \\
&& \frac{S_{BH}}{V_3}= \frac{L^3 e^{3 A(z_h)}}{4 G_5 z_{h}^{3}}. 
\label{Htemp}
\end{eqnarray}
where $V_3$ is the volume of the three-dimensional plane. Another solution which corresponds to thermal-AdS (without horizon) can be obtained by taking the limit $z_h \rightarrow \infty$, i.e.~$g(z)=1$. This thermal-AdS solution goes asymptotically to AdS at the boundary $z=0$, however it can have, depending on the scale factor $A(z)$, a non-trivial structure in the bulk.\\

It is important to mention that the form of $f(z)$ is also arbitrary and we chose $f(z)$ as in eq.~(\ref{metsolution1}) to match the holographic results for the boundary gauge theory with real QCD. For example, using $f(z)=e^{c z^2 -A(z)}$ it is easy to show, by studying the spectrum of the linear fluctuations of the gauge field, that the meson mass spectrum of the boundary gauge theory lies on a linear Regge trajectory, as dictated by QCD phenomenology. Similarly, the magnitude of parameter $c=1.16 \ \text{GeV}^2$ is fixed by comparing the holographic meson mass spectrum to that of lowest lying (heavy) meson states. From the mathematical viewpoint, our models are also consistent, for the choices of scale functions $A(z)$ that we will make in following sections, with the requirement that the argument of the root defining $\phi'(z)$, see eq.~\eqref{metsolution1}, is positive.

\section{Holographic entanglement entropy}
In this section we will derive the relevant expressions of the holographic entanglement entropy using the prescription of \cite{Ryu:2006bv}. According to this prescription, the entanglement entropy of the subsystem $B$ is given by the area of the minimal surface $\gamma_B$ which extends from the asymptotic boundary into the bulk and which shares its boundary $\partial B$ with that of the subsystem $B$. This holographic prescription therefore geometrizes the notion of entanglement entropy, as encoded in
\begin{eqnarray}
S^{EE}=\frac{\text{Area}(\gamma_B)}{4 G_5}\,.
\label{RT}
\end{eqnarray}
We work here in the Einstein frame. We checked that our results are unchanged when we switch to the string frame, given that in the latter case an extra dilaton-dependent exponential prefactor is to be added to the prescription \eqref{RT}, as in \cite{Klebanov0709,Ryu:2006ef}.\\

To calculate the entanglement entropy in our EMD  model, we consider as subsystem the strip of length $\ell$. In particular, we consider the domain $-\ell/2\leq y_1 \leq \ell/2$, $0\leq y_2 \leq L_{y_2}$ and $0\leq y_3 \leq L_{y_3}$ to define the strip geometry on the boundary. The parametrization $z=z(y_1)$ leads to the following expression,
\begin{eqnarray}
S^{EE}=\frac{L_{y_2} L_{y_3} L^3}{4 G_5} \int dy_1 \ \frac{e^{3 A(z)}}{z^3}\sqrt{1+\frac{z'^2}{g(z)}}\,.
\label{SEE}
\end{eqnarray}
There are two surfaces that are local minima of eq.~(\ref{SEE}) for the strip subsystem: a connected and a disconnected surface. Let us first calculate the entanglement entropy for the connected surface. As the corresponding Lagrangian $\mathcal{L}$ of eq.~(\ref{SEE}) does not directly depend on $y_1$, the ``Hamiltonian'' $\mathcal{H}$ is conserved,
$\frac{\p}{\p y_1}\mathcal{H}=\frac{\p}{\p y_1}[z'\frac{\delta \mathcal{L}}{\delta z'}-\mathcal{L}]=0$. This leads to the following expression,
\begin{eqnarray}
\frac{e^{3 A(z)}}{z^3\sqrt{1+\frac{z'^2}{g(z)}}}=\frac{e^{3 A(z_{*})}}{z_{*}^3}\,.
\label{minimal}
\end{eqnarray}
where $z_*$ is the turning point of the minimal area surface at which $z'(y_1)|_{z=z_*}=0$. Substituting eq.~(\ref{minimal}) into eq.~(\ref{SEE}), we get the following expression for the entanglement entropy,
\begin{eqnarray}
S^{EE}_{con}=\frac{L_{y_2} L_{y_3} L^3}{2 G_5} \int_{0}^{z_*} dz \ \frac{z_{*}^3}{z^3} \frac{e^{3 A(z)-3 A(z_*)}}{\sqrt{g(z)[z_{*}^6 e^{-6A(z_*)}-z^6 e^{-6A(z)}]}}
\label{SEEcon}
\end{eqnarray}
where $z_*$ is related to the strip length $\ell$ in the following way
\begin{eqnarray}
\ell=2\int_{0}^{z_*} dz \ \frac{z^3 e^{-3 A(z)}}{\sqrt{g(z)[z_{*}^6 e^{-6A(z_*)}-z^6 e^{-6A(z)}]}}\,.
\label{lengthSEEcon}
\end{eqnarray}
On the other hand, for the disconnected surface, we get the following expression for the entanglement entropy
\begin{eqnarray}
S^{EE}_{discon}=\frac{L_{y_2} L_{y_3} L^3}{2 G_5} \biggl[ \int_{0}^{z_d} dz \ \frac{e^{3 A(z)}}{z^3\sqrt{g(z)}} + \frac{e^{3 A(z_d)}}{2 z_{d}^3} \ell \biggr]
\label{SEEdiscon}
\end{eqnarray}
where $z_d=\infty$ or $z_d=z_h$, depending on whether the background geometry is thermal-AdS (or an AdS black hole). It is important to mention that the second term in the above equation, which comes from  the surface contribution along the horizon, is zero for the thermal-AdS background. Therefore, for the thermal-AdS background the entanglement entropy of the disconnected surface is actually independent of $\ell$. However for the AdS black hole background this second term gives a non-zero contribution to the entanglement entropy.

\section{The standard confined/deconfined phases: thermal-AdS vs.~black hole}
As in \cite{Dudal:2017max}, we first consider the following simple form of $A(z)$,
\begin{eqnarray}
A(z)=A_{1}(z)=- \bar{a} z^2.
\label{Aansatz1}
\end{eqnarray}
It is easy to see that $A_{1}(z)\rightarrow 0$ at the boundary $z=0$, asserting that spacetime asymptotes to AdS. Note that near the boundary,
\begin{eqnarray}
& &V(z)|_{z\rightarrow 0}=-\frac{12}{L^2}+\frac{\Delta(\Delta-4)}{2}\phi^2(z)+\ldots, \nonumber \\
& & V(z)|_{z\rightarrow 0}=2 \Lambda + \frac{m^2 \phi^2}{2}+\ldots
\label{Vcase1exp}
\end{eqnarray}
where $m^2=\Delta(\Delta-4)$ with $\Delta=3$, satisfying the well known relation of the gauge/gravity duality.
The parameter $\bar{a}= 0.145$ in eq.~(\ref{Aansatz1}) is obtained by demanding the critical temperature $T_c$ of the thermal-AdS/black hole (or the dual confinement/deconfinement) phase transition to be around $270 \ \text{MeV}$ at zero chemical potential, following the large $N$ lattice estimate of \cite{Lucini:2003zr}.

\subsection{Black hole thermodynamics}
\begin{figure}[h!]
\begin{minipage}[b]{0.5\linewidth}
\centering
\includegraphics[width=2.8in,height=2.3in]{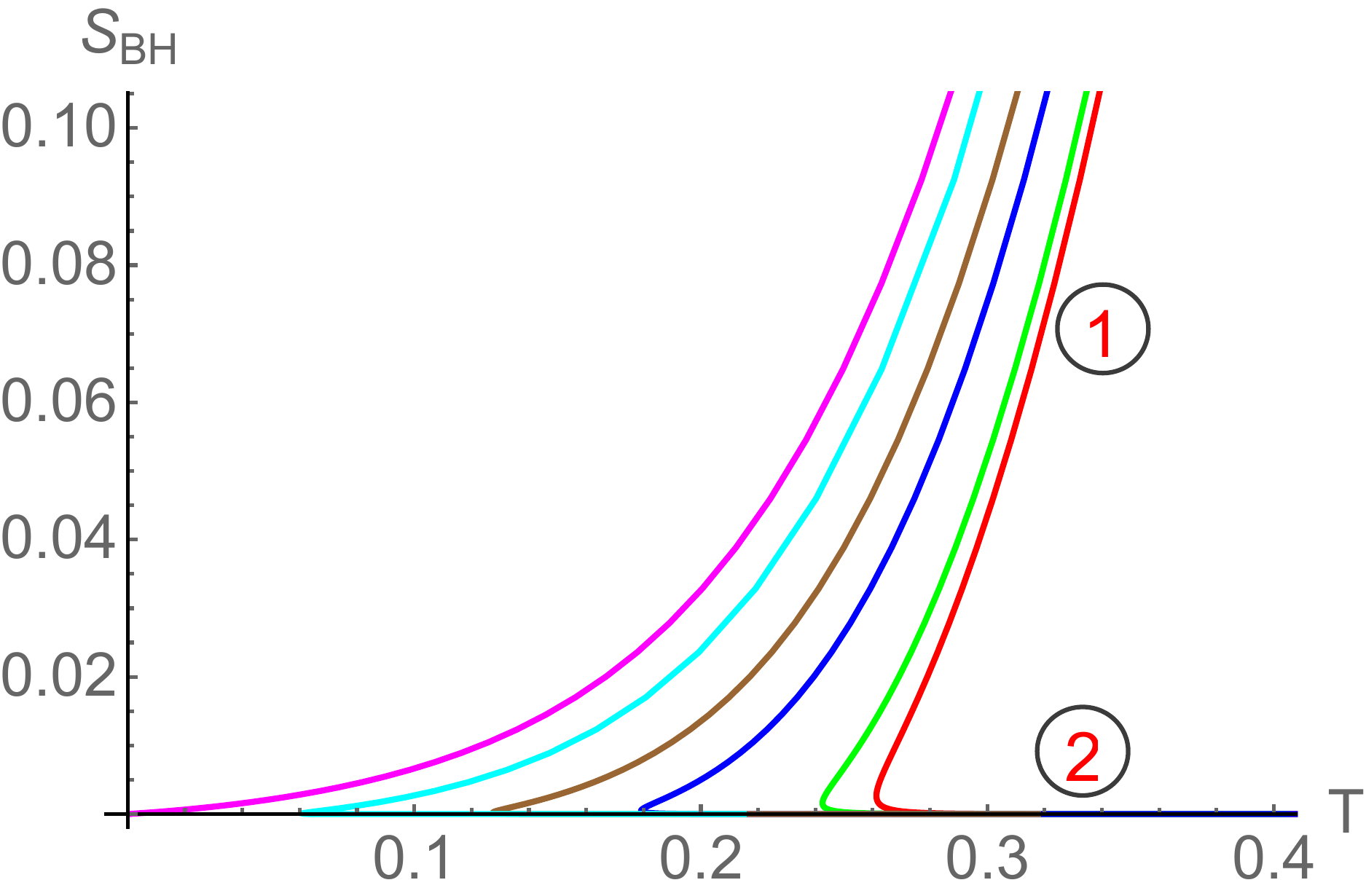}
\caption{ \small $S_{BH}$ as a function of $T$ for various values of the chemical potential $\mu$. Here red, green, blue, brown, cyan and magenta curves correspond to $\mu=0$, $0.2$, $0.4$, $0.5$, $0.6$ and $0.673$ respectively. In units \text{GeV}.}
\label{TvsSBHvsMucase1}
\end{minipage}
\hspace{0.4cm}
\begin{minipage}[b]{0.5\linewidth}
\centering
\includegraphics[width=2.8in,height=2.3in]{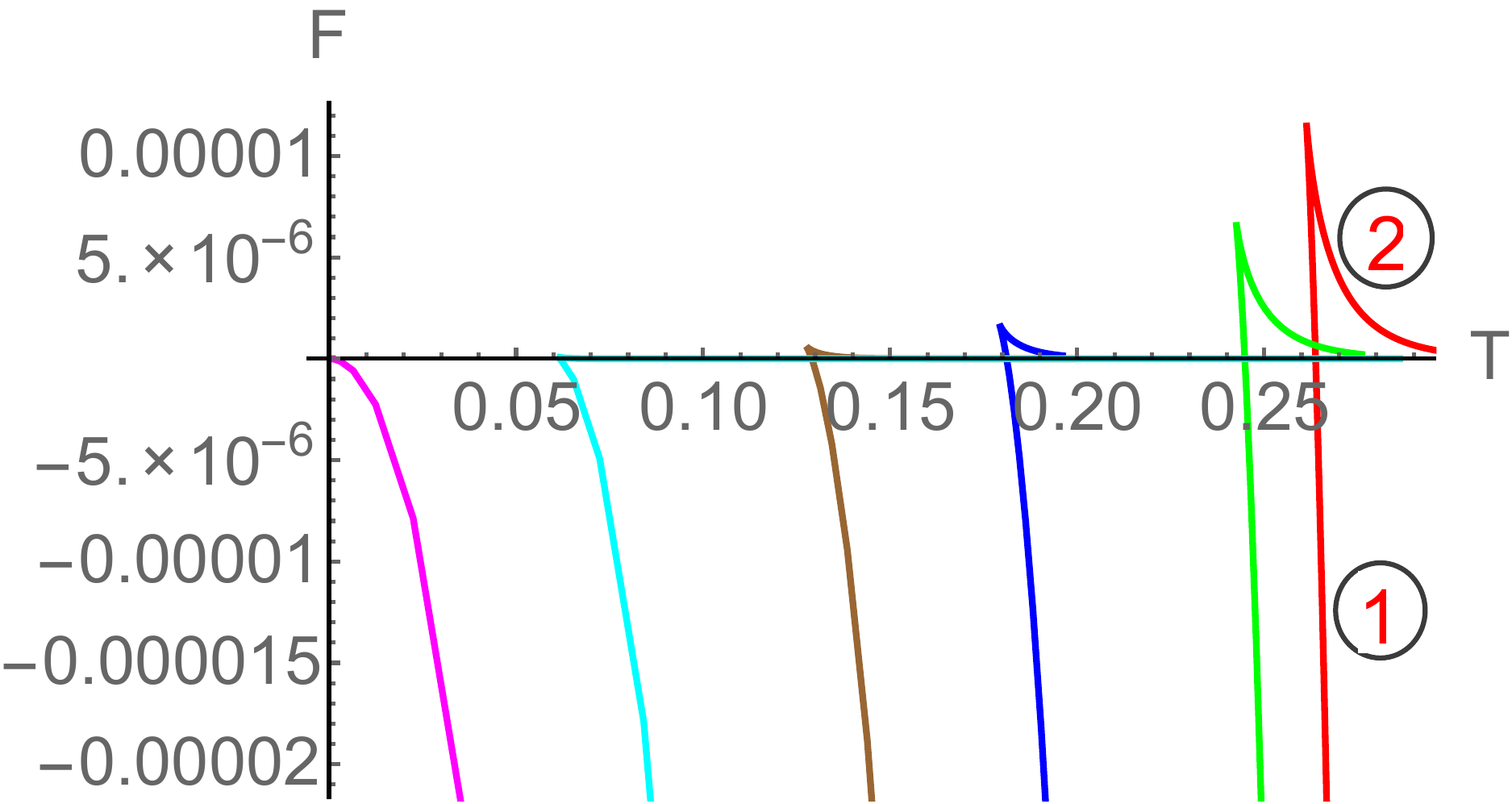}
\caption{\small $F$ as a function of $T$ for various values of the chemical potential $\mu$. Here red, green, blue, brown, cyan and magenta curves correspond to $\mu=0$, $0.2$, $0.4$, $0.5$, $0.6$ and $0.673$ respectively. In units \text{GeV}.}
\label{TvsFBHvsMucase1}
\end{minipage}
\end{figure}
The thermodynamics of the gravity solution with eq.~(\ref{Aansatz1}) is shown in Figures~\ref{TvsSBHvsMucase1} and \ref{TvsFBHvsMucase1}. For small values of $\mu$, we find two black hole solutions at each temperature: a large black hole solution (corresponding to small $z_h$) which is stable and a small black hole solution (corresponding to large $z_h$) which is unstable. These two black hole solutions are shown in Figure~\ref{TvsSBHvsMucase1} in the $(T, S_{BH})$ plane. The branch with positive slope, for which the entropy increases with temperature, is stable whereas the branch with negative slope, for which the entropy decreases with temperature, is unstable. These stable and unstable black hole solutions are marked by $\circled{1}$ and $\circled{2}$. Furthermore, the black hole solution only exists above a minimum temperature $T_{min}$, suggesting a phase transition to thermal-AdS as we decrease the temperature. Indeed as shown in Figure~\ref{TvsFBHvsMucase1}, the free energy, which is normalized with respect to thermal-AdS, changes its sign as the temperature decreases. The sign change takes place at $T_{c} > T_{min}$, indicating a first order phase transition from AdS black hole to thermal-AdS as the temperature decreases. This is the famous Hawking--Page phase transition. \\

We also note that for higher values of $\mu$ the magnitude of the negative slope branch starts decreasing and becomes positive after some critical $\mu_c$. In particular, at $\mu_c$ the unstable branch disappears and we have a single black hole solution which remains stable at all temperatures. This is indicated by a magenta line in Figures~\ref{TvsSBHvsMucase1} and \ref{TvsFBHvsMucase1}. For this model, we get $\mu_c=0.673 \ \text{GeV}$. We find that the critical temperature decreases with $\mu$ and the first order phase transition stops at $\mu_c$. A determination of the QCD critical point, if existing, in the $(T,\mu)$ plane is notoriously difficult \cite{Ratti:2018ksb,deForcrand:2002hgr,Brewer:2018abr}, a reasonable estimate is a few hundred MeV, so we see our estimate lies in the same ballpark, as well as the one we will find with our second choice for the form factor $A(z)$.
\\

In \cite{Dudal:2017max}, we used this Hawking-Page phase transition on the gravity side to determine the confinement/deconfinement phase transition on the dual boundary side. By confinement, we mean here a phase for which the free energy of the probe quark-antiquark pair varies linearly with respect to their separation length, leading to an area law for the Wilson loop, while the Polyakov loop expectation value vanishes. We demonstrated that the boundary theory dual to the thermal-AdS phase does indeed show these properties whereas, the one dual to the AdS black hole phase does not. Moreover, as shown above, the critical temperature of the dual confinement/deconfinement phase transition decreases with chemical potential which is again in line with realistic QCD. This model therefore provided us with a more realistic holographic realization of QCD, compared to the soft wall model, for the confinement/deconfinement phase structure, this in particular because we have included the backreaction of the dilaton field in a self-consistent form from the beginning. It is therefore of interest to investigate the entanglement entropy in this more trustable holographic QCD model and see whether it can again correctly distinguish the nature of confined and deconfined phases.

\subsection{Holographic entanglement entropy}
Our aim in this subsection is to calculate the entanglement entropy holographically in the above constructed confined/deconfined phases. The necessary formulas have already been worked out in the previous section and here we present our numerical results.\\
\begin{figure}[h!]
\begin{minipage}[b]{0.5\linewidth}
\centering
\includegraphics[width=2.8in,height=2.3in]{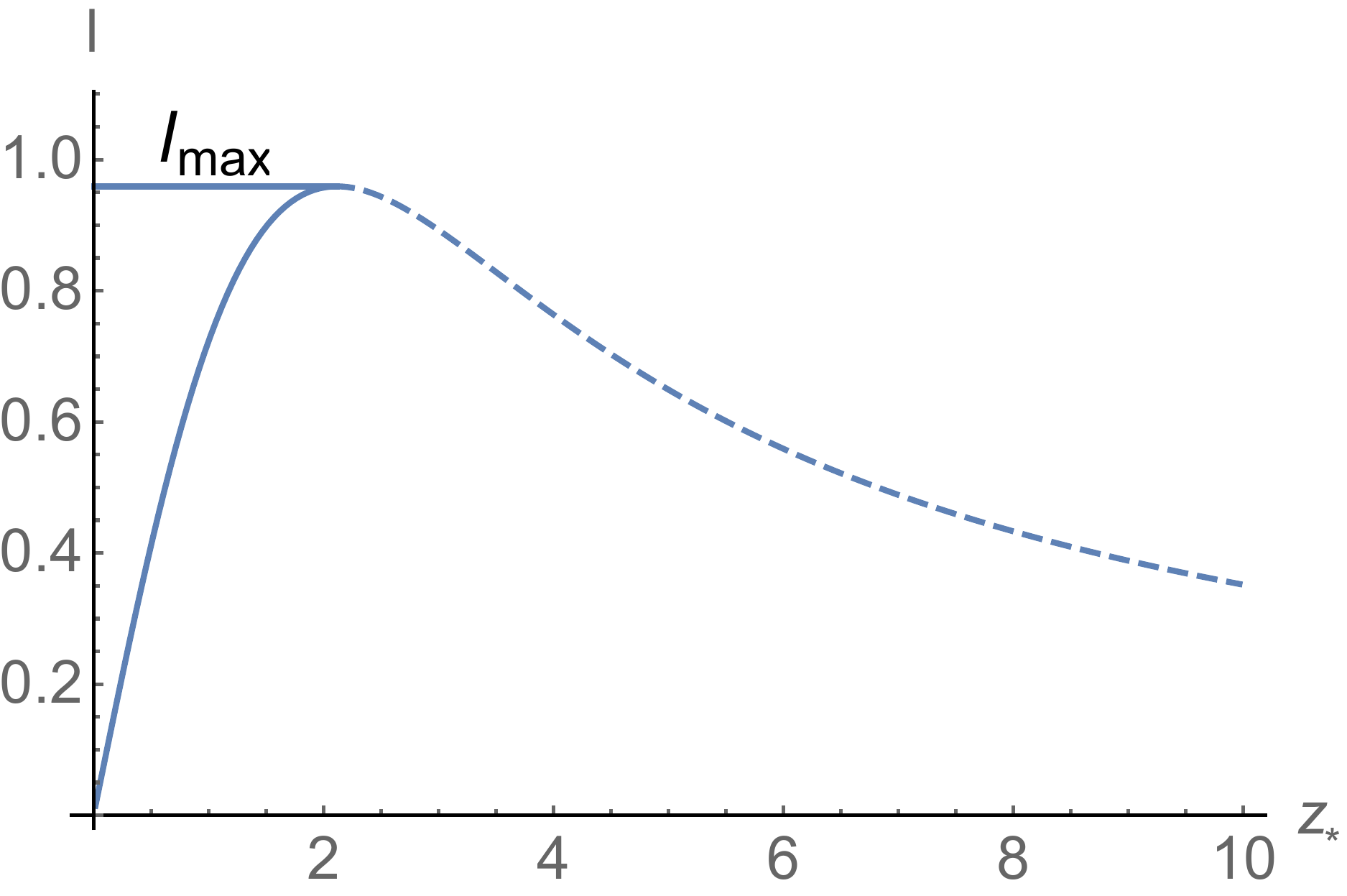}
\caption{ \small $\ell$ as a function of $z_*$ in the thermal-AdS background. In units \text{GeV}.}
\label{zsvslAdScase1}
\end{minipage}
\hspace{0.4cm}
\begin{minipage}[b]{0.5\linewidth}
\centering
\includegraphics[width=2.8in,height=2.3in]{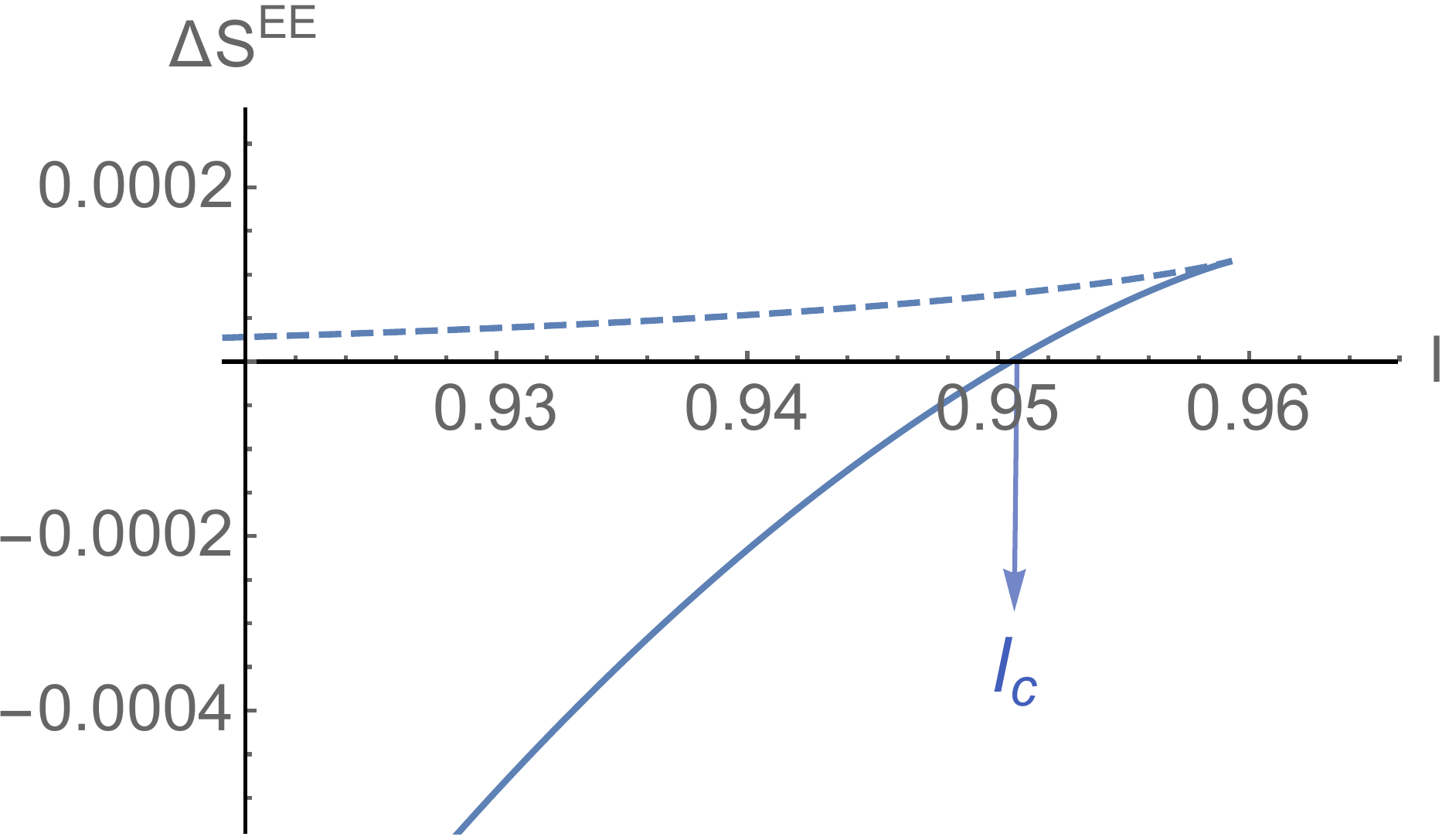}
\caption{\small $\Delta S^{EE}=S^{EE}_{con}-S^{EE}_{discon}$ as a function of $\ell$ in the thermal-AdS background. In units \text{GeV}.}
\label{lvsSEEAdScase1}
\end{minipage}
\end{figure}

Let us first discuss the results in the thermal-AdS background. In Figure~\ref{zsvslAdScase1}, we have shown the variation of subsystem size $\ell$ with respect to the turning point $z_*$ of the connected entangling surface. We find that there exists a maximum length $\ell_{max}$ above which the connected entangling surface does not exist and only the disconnected entangling surface remains. Moreover, below $\ell_{max}$ there are two solutions of eq.~(\ref{lengthSEEcon}) for a given $\ell$ that minimizes the entanglement entropy expression of eq.~(\ref{SEEcon}). The one for small $z_*$ (solid line) corresponds to an actual local minimum whereas the one for large $z_*$ (dashed line) corresponds to a saddle point.\\

In Figure~\ref{lvsSEEAdScase1}, the difference between the connected and disconnected entanglement entropies $\Delta S^{EE}=S^{EE}_{con}-S^{EE}_{discon}$  is shown. Here solid and dash lines are again corresponding to small and large $z_*$ solutions. We see that the solution which corresponds to small $z_*$ always has a lower entanglement entropy than the large $z_*$ solution, justifying our earlier statement that the former solution is the true minimum. Interestingly, we find that $\Delta S^{EE}$ changes sign from negative to positive value as we increase the subsystem size. $S^{EE}_{con}$ has a lower entanglement entropy for small $\ell$ whereas $S^{EE}_{discon}$ has a lower entanglement entropy for large $\ell$. This implies a phase transition from connected to disconnected entangling surfaces as we increase the subsystem size. The length at which this phase transition occur defines an $\ell_{c}(<\ell_{max}$). For $c=1.16$, we find\footnote{In general $\ell_{c}$ depends on the parameter $c$ and can have a different magnitude for different $c$.}
\begin{equation}\label{lcr}
\ell_{c}\simeq 0.95 \ \text{GeV}^{-1}.
\end{equation}
As mentioned before, for the thermal-AdS background the minimal area of the disconnected surface is independent of $\ell$, implying that the corresponding entanglement entropy  is as well. Consequently, for large $\ell(>\ell_{c})$, the entanglement entropy of the strip subsystem becomes independent of $\ell$.  These results therefore can be summarized in the following way,
\begin{eqnarray}
\frac{\partial S^{EE}}{\partial \ell} &\propto &\frac{1}{G_N} = \mathcal{O}(N^2)\quad\text{for}\quad \ell < \ell_{c}\,, \nonumber \\
&\propto& \frac{1}{G_{N}^{0}} = \mathcal{O}(N^0)\quad\text{for}\quad \ell > \ell_{c}
\end{eqnarray}
where the supposedly large number $N$ corresponds to the number of colors. \\

This kind of phase transition between two entangling surfaces was first observed in \cite{Klebanov0709} for top-down models in gauge/gravity duality and it was suggested to be a characteristic feature of confining gauge theories. In particular, the entanglement entropy was shown to scale as $N^2$ for small $\ell$ and as $N^0$ for large $\ell$. This naturally led to the interpretation of the subsystem size $\ell$ as the inverse temperature ``$T_{c}\propto \frac{1}{\ell_{c}}$'' \footnote{$T_{c}$ here should not be confused with the $T_{c}$ associated with the Hawking-Page phase transition.}. Indeed, above the deconfinement critical temperature the deconfined colored gluon degrees of freedom, which are in an adjoint representation, count as order $\mathcal{O}(N^2)$ whereas below this critical temperature the color-neutral confined degrees of freedom count as order $\mathcal{O}(N^0)$, the same counting as suggested by the entanglement entropy. Moreover, for a small system we expect perturbative UV fluctuations to be the dominant contribution, which cannot be the source of confinement.\\

We thus find a similar result for the entanglement entropy, however now in a self-consistent bottom-up confining model. Importantly, as we will show shortly, the connected to disconnected surface transition appears only in the confined phase and no such transition occurs in the deconfined or specious-confined phases, albeit that in the latter, another transition occurs. In any case, our analysis further substantiates the claim made by \cite{Klebanov0709} for the entanglement entropy as a probe to diagnose confinement.
 Let us briefly comment on the size of $\ell_c$, which is roughly compatible with the size of the (spherical) MIT bag model \cite{Chodos:1974je,Signal:1988vf} used to model nucleons. A more precise characterization to connect the size of the entangling surface to (a) physical QCD scale(s) would be to consider either a sphere (to model the bag) or finite cylinder (to model the physical QCD flux tube between heavy static quarks, which has a finite width as well, see e.g.~\cite{Cardoso:2013lla}). Inside the bag or flux tube, we expect the QCD vacuum to be dominated by perturbative, short-range fluctuations, while outside the non-perturbative dynamics becomes dominantly important. A popular visualization of the QCD vacuum is that of a dual type II superconductor, according to which the string tension inside the flux tube should vanish, so inside there is actually deconfinement, coming with evidence from lattice QCD simulations \cite{DiGiacomo:1990hc,DiGiacomo:1996pp}. We might thus expect the entanglement entropy to be sensitive to the size of the bag or flux tube cylinder. In fact, the entanglement entropy could help to shed light on determining from first principles the size of the bag or flux tube or to better understand the origin of this dual superconductor vacuum scenario, besides the lattice picture. It is interesting to mention here that the entanglement entropy has been used explicitly in the context of superconductors to probe the different phases \cite{Albash:2012pd,Dey:2014voa,Cai:2012sk}.  Investigating the just sketched bag or flux tube phenomenology deeper will be a highly non-trivial exercise, but we hope to come back to this issue in future work, using the techniques of e.g.~\cite{deBoer:2011wk} to compute holographic entanglement entropies for more sophisticated, physical geometries.\\

Moreover, the non-analytic behavior in the structure of entanglement entropy that we have discussed above has also been found in SU(2) gauge theory using lattice calculations \cite{Buividovich:2008kq}. Such non-analyticity in the nature of entanglement entropy therefore seems to be a common feature of all confining gauge theories. Our holographic prediction for the length scale ($\ell_{c} \sim 0.2 \ \text{fm}$) at which the non-analyticity in the entanglement entropy appears is of same order as predicted by the lattice simulations ($\ell_c \sim 0.5 \ \text{fm}$) \cite{Buividovich:2008kq,Itou:2015cyu}. This interesting result again lends a strong support for the general belief that the idea of gauge/gravity duality can provide testable predictions for QCD-like gauge theories. As such, as we will also see in next section, the introduction of temperature in the specious confined phase leads to a phase diagram in the $(T,\ell_{c})$ plane, which interestingly again compares qualitatively well with the lattice QCD intuition put forward in \cite{Buividovich:2008kq}.
\\

\begin{figure}[h!]
\begin{minipage}[b]{0.5\linewidth}
\centering
\includegraphics[width=2.8in,height=2.3in]{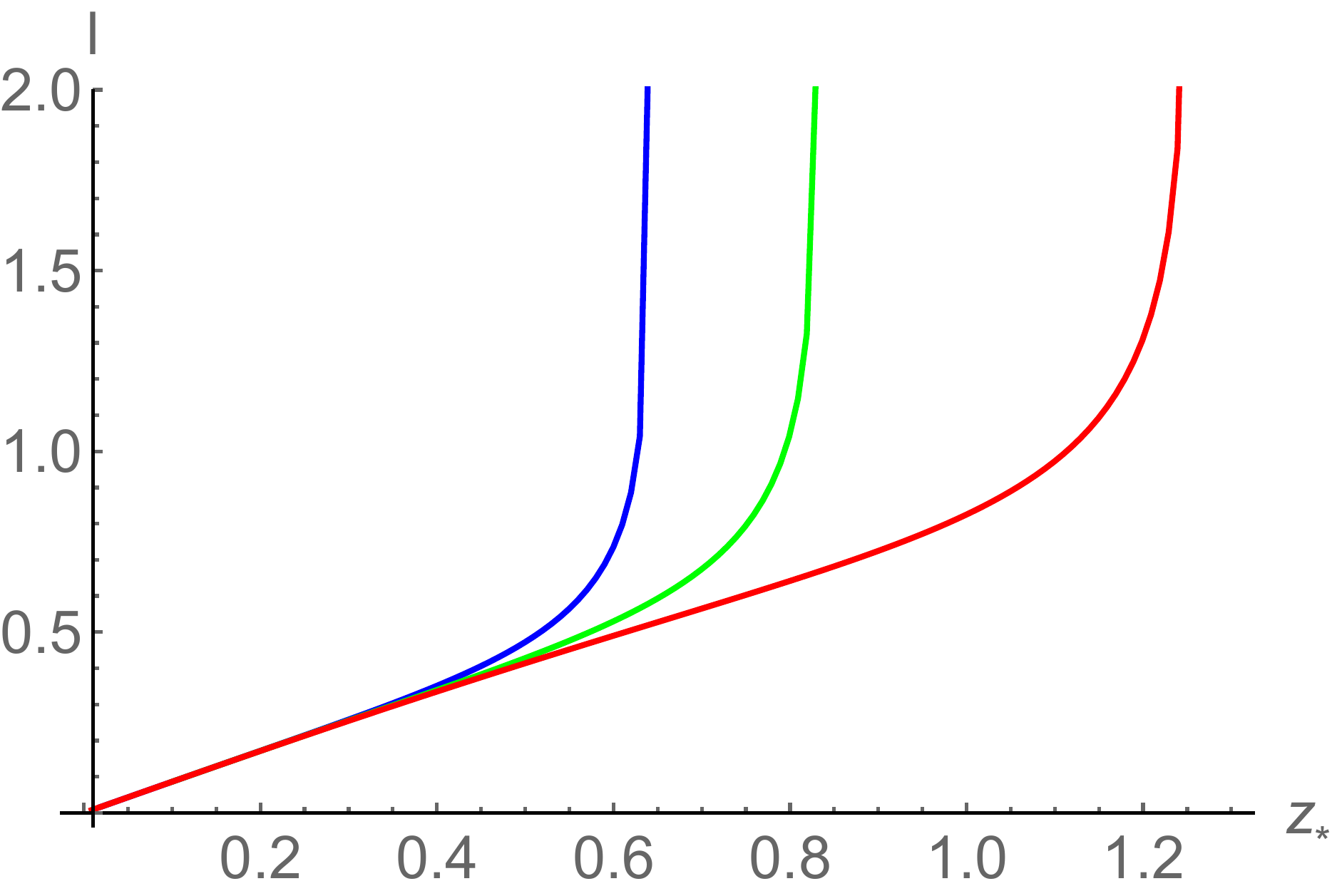}
\caption{ \small $\ell$ as a function of $z_*$ in the AdS black hole background. Here $\mu=0$ and red, green and blue curves correspond to $T/T_{c}=1.2$, $1.6$ and $2.0$ respectively. In units \text{GeV}.}
\label{zsvslAdSBHMu0case1}
\end{minipage}
\hspace{0.4cm}
\begin{minipage}[b]{0.5\linewidth}
\centering
\includegraphics[width=2.8in,height=2.3in]{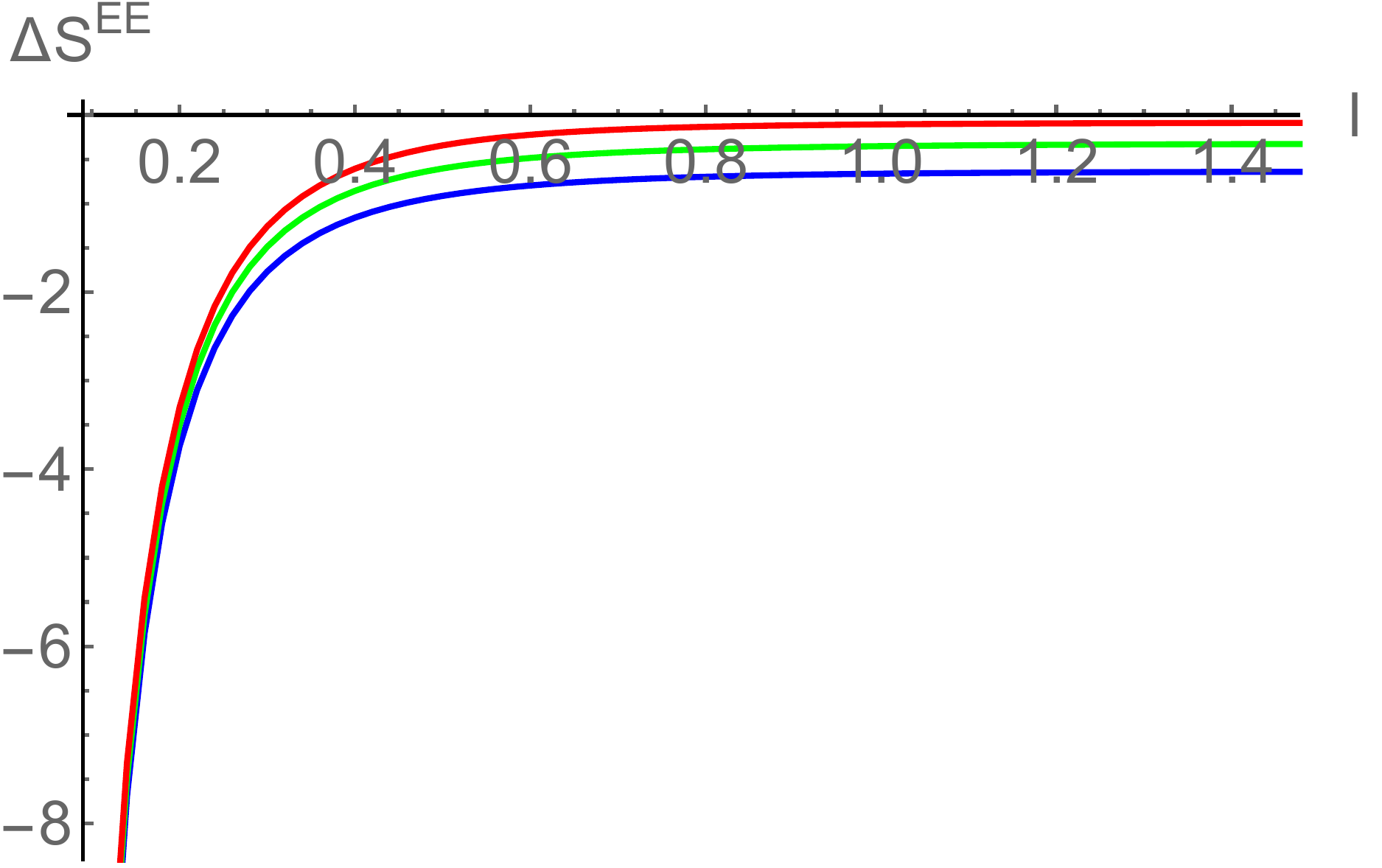}
\caption{\small $\Delta S^{EE}=S^{EE}_{con}-S^{EE}_{discon}$ as a function of $\ell$ in the AdS black hole background. Here $\mu=0$ and red, green and blue curves correspond to $T/T_{c}=1.2$, $1.6$ and $2.0$ respectively. In units \text{GeV}.}
\label{lvsSEEAdSBhMu0case1}
\end{minipage}
\end{figure}

Having discussed the holographic entanglement entropy with the thermal-AdS background, we now move on to discuss it with the AdS black hole background. The results are shown in Figures~\ref{zsvslAdSBHMu0case1} and \ref{lvsSEEAdSBhMu0case1}.\\

With the AdS black hole background, the situation is quite different. In particular, no $\ell_{max}$ does appear and the connected entangling surface continues to exist for any $\ell$. This is shown in Figure~\ref{zsvslAdSBHMu0case1}, where a one to one relation between $z_*$ and $\ell$ can be explicitly seen. It further indicates that for higher and higher values of $\ell$, the turning point of the connected entangling surface moves closer towards the horizon $z_h$.  Moreover, we find that no $\ell_{c}$  exists as well. In particular, no phase transition from a connected to disconnected entangling surface shows up as we increase the subsystem size. This is shown in Figure~\ref{lvsSEEAdSBhMu0case1} where we see that
$\Delta S^{EE} \leq 0$, indicating $S^{EE}_{con}\leq S^{EE}_{discon}$. It is important to mention that it is the second term of eq.~(\ref{SEEdiscon}), which is absent in the thermal-AdS background, that makes $S^{EE}_{discon}\geq S^{EE}_{con}$. The equality sign only is realized when the subsystem size approaches the full system size, i.e.~when $\ell \rightarrow \infty$. It is interesting to note that in this limit, we get
\begin{eqnarray}
S^{EE}_{con} =S^{EE}_{discon} = \frac{L_{y_2} L_{y_3} L^3}{4 G_5} \frac{e^{3 A(z_h)}}{ z_{h}^3} \ell
\end{eqnarray}
which is nothing but the Bekenstein--Hawking entropy of the black hole. Of course this is expected from a general property of the entanglement entropy, which states that at finite temperature the entanglement entropy approaches the thermal entropy when the size of the subsystem approaches its full system size.\\
\begin{figure}[h!]
\begin{minipage}[b]{0.5\linewidth}
\centering
\includegraphics[width=2.8in,height=2.3in]{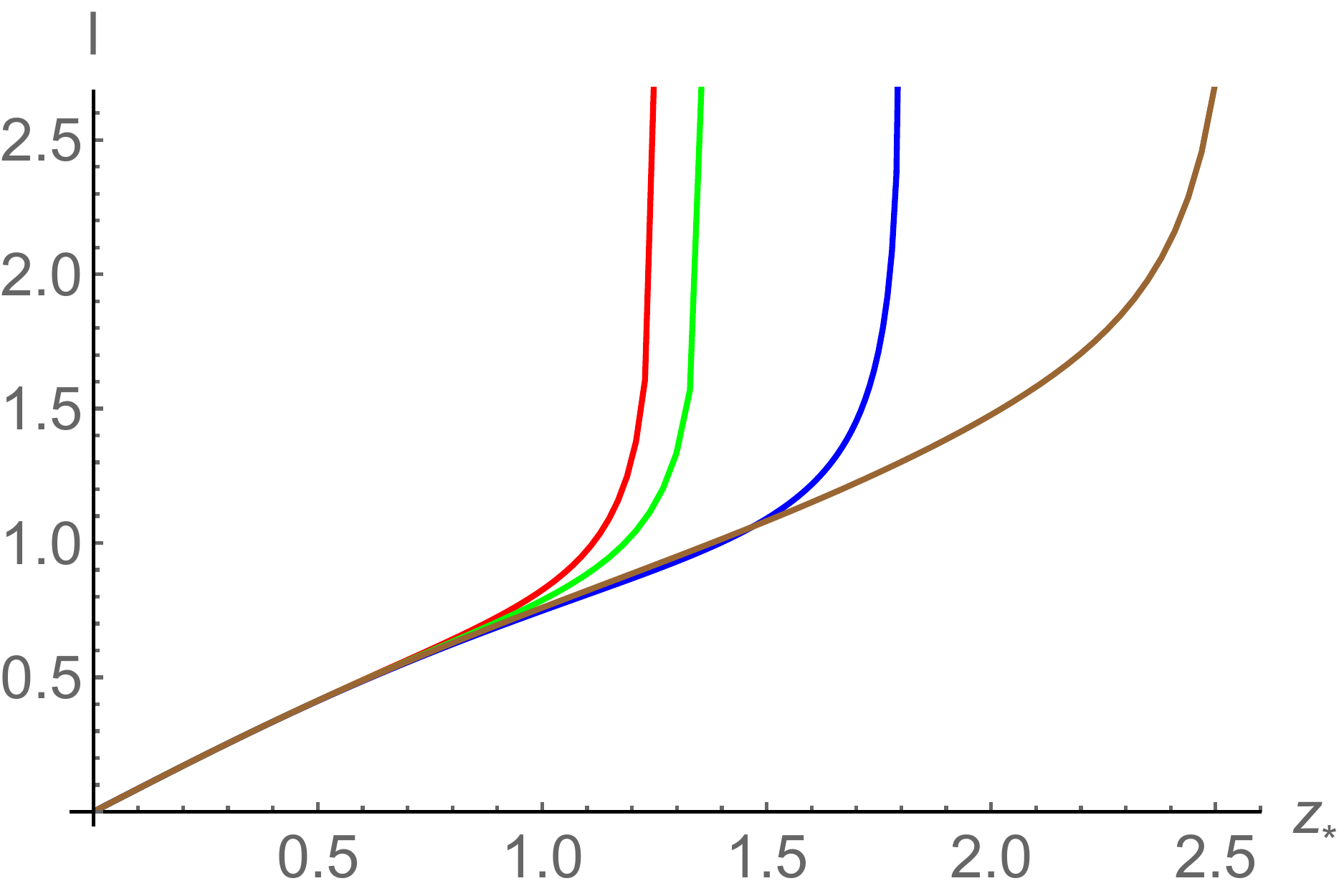}
\caption{ \small $\ell$ as a function of $z_*$ in the deconfined phase. Here $T/T_{c}=1.2$ and red, green, blue and brown curves correspond to $\mu=0$, $0.2$, $0.4$ and $0.6$ respectively. In units \text{GeV}.}
\label{zsvslvsMuAdSBHcase1}
\end{minipage}
\hspace{0.4cm}
\begin{minipage}[b]{0.5\linewidth}
\centering
\includegraphics[width=2.8in,height=2.3in]{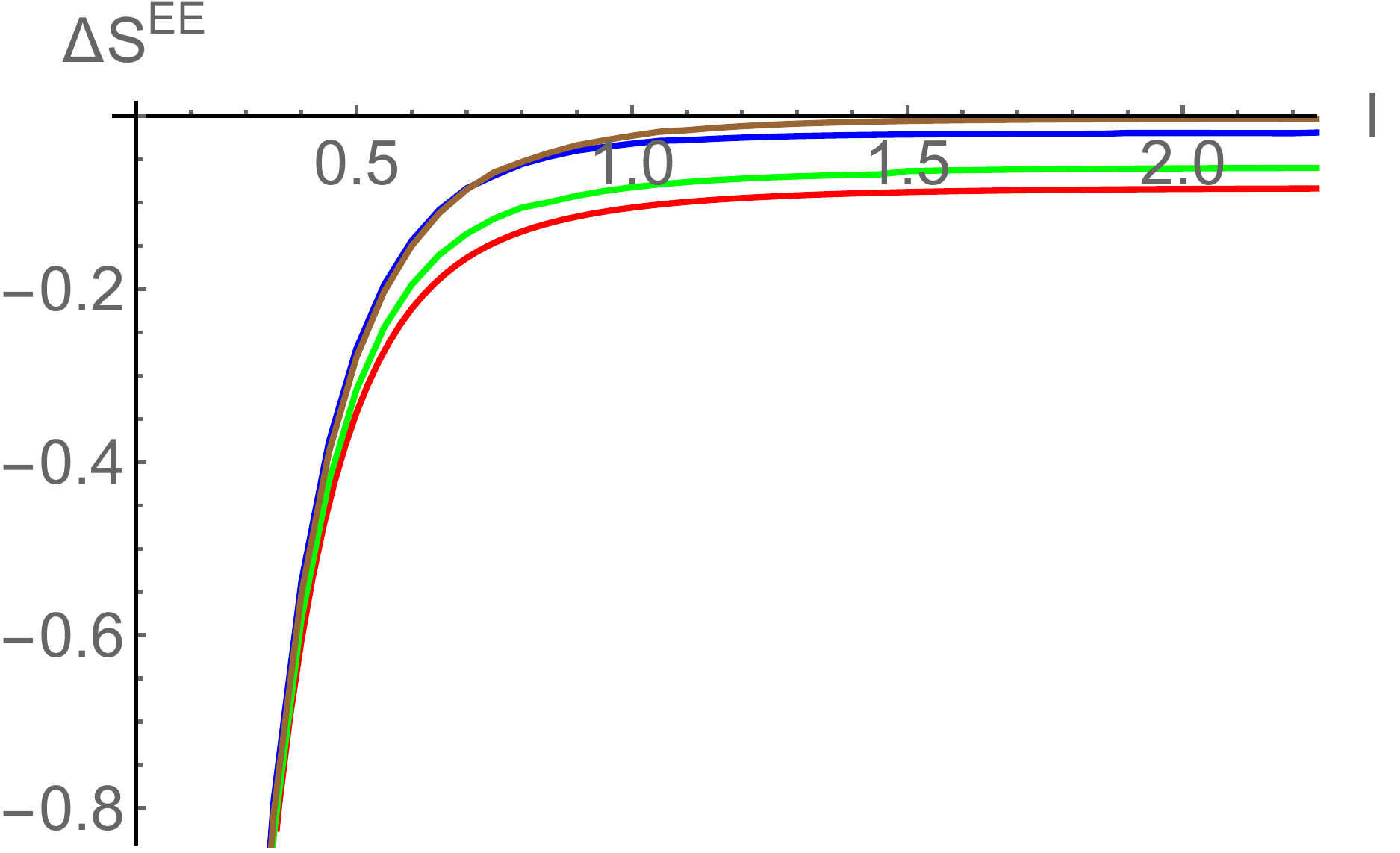}
\caption{\small $\Delta S^{EE}=S^{EE}_{con}-S^{EE}_{discon}$ as a function of $\ell$ in the in the deconfined phase. Here $T/T_{c}=1.2$ and red, green and blue curves correspond to $\mu=0$, $0.2$, $0.4$ and $0.6$ respectively. In units \text{GeV}.}
\label{lvsSEEvsMuAdSBhcase1}
\end{minipage}
\end{figure}

Importantly, for the AdS black hole background we always have,
\begin{eqnarray}
\frac{\partial S^{EE}}{\partial \ell} \propto \frac{1}{G_N} = \mathcal{O}(N^2)\,.
\end{eqnarray}
These results for the deconfined phase are present for other values of chemical potential as well. The results at $T=1.2 \ T_{c}$ are shown in Figures~\ref{zsvslvsMuAdSBHcase1} and \ref{lvsSEEvsMuAdSBhcase1}.  The connected surface again exists for all $\ell$ whose corresponding entanglement entropy is smaller than its disconnected counterpart, indicating no phase transition between them as the subsystem size varies.
\\

We now move on to discuss the entanglement entropy as function of temperature and chemical potential, which is of paramount relevance for the characterization and understanding of the holographic
QCD phase diagram. A complete study of entanglement entropy in the parameter space of $T$ and $\mu$ will serve us two purposes. First it will help us to understand the thermodynamic aspects of the entanglement entropy and its connection with the black hole phase transition. Indeed, as mentioned before, our gravity background undergoes a phase transition from thermal-AdS to AdS black hole as we vary the Hawking temperature. Since the Ryu--Takayanagi prescription geometrizes the definition of entanglement entropy, whose corresponding minimal area surface propagates from the asymptotic boundary into the bulk, it is a natural question to ask whether the holographic entanglement entropy sees the signature of this black hole phase transition. Second, since the back hole phase transition corresponds to the confinement/deconfinement phase transition in the dual boundary theory, it will also help to probe the QCD phase diagram holographically from the entanglement entropy perspective.

\begin{figure}[h!]
\centering
\includegraphics[width=2.8in,height=2.3in]{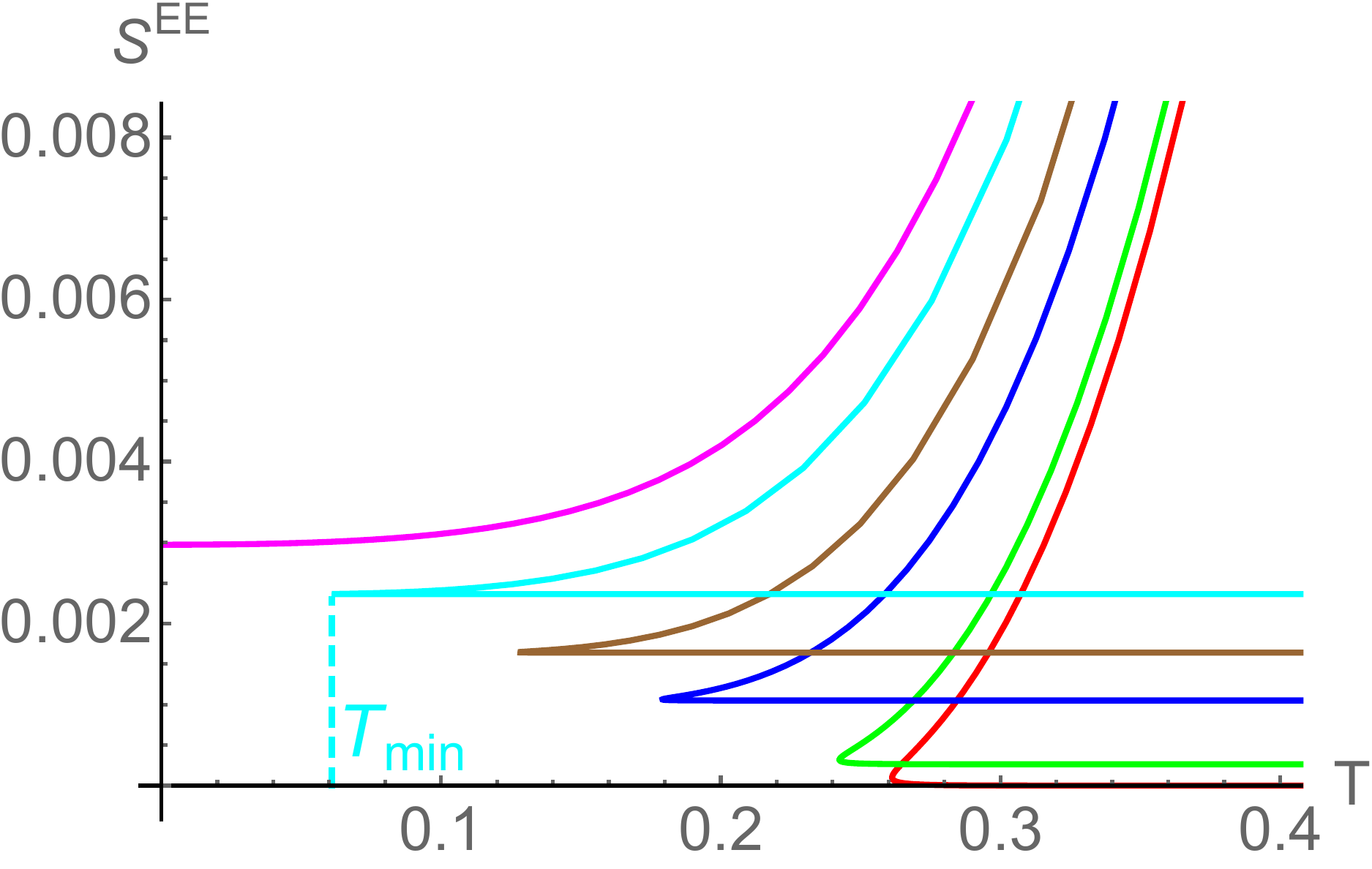}
\caption{ \small $\Delta S^{EE}=S^{EE}_{AdS-BH}-S^{EE}_{Thermal \ AdS}$ as a function of $T$. Here $\ell=0.2$ and red, green, blue, brown, cyan and magenta curves correspond to $\mu=0.0$, $0.2$, $0.4$, $0.5$, $0.6$ and $0.673$ respectively.}
\label{TvsdelSEEvsMucase1}
\end{figure}

Our result is shown in Figure~\ref{TvsdelSEEvsMucase1}, which displays $\Delta S^{EE}=S^{EE}_{AdS-BH}-S^{EE}_{Thermal-AdS}$ as a function of Hawking temperature for various values of $\mu$. Here $S^{EE}_{Thermal-AdS}$, which is independent of temperature and constant for a given strip length $\ell$, is used to subtract the divergence from $S^{EE}_{AdS-BH}$.
First, we observe a striking similarity in the structure of entanglement entropy with the Bekenstein--Hawking thermal entropy of the black hole (shown in Figure~\ref{TvsSBHvsMucase1}). In particular, there are again two branches in the entanglement entropy for small values of $\mu$. The branch with positive slope corresponds to the stable solution whereas the branch with negative slope corresponds to the unstable solution. The unstable solution does not exist above the same $\mu_c$ as before and there is only branch which is stable at all temperatures (shown by the magenta line).\\

At first sight, the similarity between the entanglement and Bekenstein--Hawking entropy (which is also the thermal entropy of the boundary QCD) does not seem very surprising. As mentioned earlier, the entanglement entropy approaches the Bekenstein--Hawking entropy for $\ell \rightarrow \infty$, and therefore it is expected that the entanglement entropy might share a few properties with the Bekenstein--Hawking entropy in the limit of a large subsystem. However, what is more surprising is that the temperature dependent behavior of the entanglement and Bekenstein--Hawking entropy remains even similar for smaller $\ell$. In Figure~\ref{TvsdelSEEvsMucase1}, we have used $\ell=0.2 \ \text{GeV}^{-1}$. However, we have checked for other values of $\ell$ as well that the essential features of Figure~\ref{TvsdelSEEvsMucase1}  remain unchanged.  With different $\ell$, only the magnitude of the entanglement entropy changes and other useful quantities such as $\mu_c$ and $T_{min}$ are unaffected. These findings are in qualitative agreement with the observations in \cite{Knaute:2017lll}.\\

In recent years, due to its above mentioned similarities with the Bekenstein--Hawking entropy, the entanglement entropy has been suggested as an efficient tool to probe even the black hole phase transitions and it has been explored in many different gravity theories \cite{Johnson:2013dka,Dey:2015ytd,Nguyen:2015wfa,Zeng:2015wtt,Zeng:2016sei,Sun:2016til,Zeng:2016aly,Liu:2017jbm,ElMoumni:2016eqh,Mo:2017oqj,ElMoumni:2018fml}. Our analysis adds further weight on this growing belief as we found similar results in a more advanced phenomenological bottom-up holographic model. Moreover, as we will show in the next section, even by taking a more complicated form of the scale actor $A(z)$ which will instead lead to a small/large black hole phase transition, the entanglement entropy continues to show similarities with the Bekenstein--Hawking entropy, capturing the essence of the phase transition.\\

An important point that we would like to point out, and which has not been emphasized much in the literature to our knowledge, is that although the entanglement entropy does seem to capture the presence of the black hole (or dual confinement/deconfinement) phase transition, it does not provide any information about the critical point $T_{c}$, especially if the transition is of first order. The $T$ vs.~$S^{EE}$ curve only emphasizes the importance of $T_{min}$, around which $S^{EE}$ starts to show double valuedness. However, the first order phase transition actually occur at $T_{c}>T_{min}$ and that cannot be obtained or pin-pointed by $S^{EE}$. Therefore, we need more information than what we get from the entanglement entropy alone to have a complete information about the phase transition. However $\mu_c$, unlike $T_{c}$, can be very well referred from the entanglement entropy, see also the next section.\\

On the other hand, as it is well known in condensed matter (and in the area of holographic superconductors as well) that the entanglement entropy generally scales differently above and below the critical point for systems which undergo a second order phase transition \cite{Osborne:2002zz,Albash:2012pd,Dey:2014voa} \footnote{Here the critical point can be achieved by varying external parameters such as the coupling constant and not just the temperature.}. In those systems, the entanglement entropy can be used to obtain information about the critical point as well.

\section{The specious-confinement/deconfinement phases: small vs.~large black hole}
In \cite {Dudal:2017max}, we constructed a novel model of holographic QCD by considering the following complicated form of $A(z)$,
\begin{eqnarray}
A(z)=A_{2}(z)=-\frac{3}{4}\ln{(a z^2+1)}+\frac{1}{2}\ln{(b z^3+1)}-\frac{3}{4}\ln{(a z^4+1)}
\label{Aansatz2}
\end{eqnarray}
where on the boundary side a close cousin of the standard confined phase, which we called the specious-confined phase, was revealed. Importantly, this specious-confined phase was shown to be also dual to a black hole configuration on the gravity side. The black hole configuration naturally led to the notion of temperature in the specious-confined phase, which in turn had important consequences as we could now study the temperature dependent properties of the specious-confined phase as well. Importantly, we showed that the specious-confined phase resembles very well the usual QCD confined phase in many ways and share many of its properties. \\

It is  again easy to see that the bulk spacetime asymptotes to AdS with the above choice of scale factor as well, by means of $A_{2}(z)\rightarrow 0$ at the boundary $z=0$. More specifically, we also have that the dilaton $\phi(z)\rightarrow0$ as $z\rightarrow0$, as expected in the conformal limit. The parameters $a=\frac{c}{9}$ and $b=\frac{5c}{16}$ are again chosen to reproduce some of real QCD results holographically. In particular, with these choices we get the specious-confined/deconfined transition temperature around 270~MeV at zero chemical potential. Further, the dilaton potential has the same asymptotic form for both $A_{1}(z)$ and $A_{2}(z)$. In particular, at the asymptotic boundary where both $A_{1}(z)$ and $A_{2}(z)$ approaches zero, we get $V(z)|_{z\rightarrow 0}=-\frac{12}{L^2}=2 \Lambda$. In general, for both choices $A_{1,2}(z)$, it can also be verified that $V(0)\geq V(z)$, i.e.~the potential is bounded from above by its UV boundary value, as prescribed by \cite{Gubser:2000nd} to have a well-defined boundary theory. In the deep UV, the strong coupling constant $\alpha_s$ goes to a constant in our case, as $\alpha_s$ is usually identified with $e^{\phi}$. If one would wish to also incorporate asymptotic freedom with a $\beta$-function resembling that of QCD, an extra term $\propto \frac{1}{\ln z}$ can be added to $A_{1,2}(z)$, see for instance \cite{Paula} or \cite{Pirner:2009gr,Galow:2009kw,He:2010ye}. In any case, we have verified that such term is not crucial to find the linear confining potential while its adding only quantitatively influences the main reported features in the confined/deconfined phases regarding the interquark potential, the entropy, the speed of sound or the entanglement entropy.

\subsection{Black hole thermodynamics}
\begin{figure}[h!]
\begin{minipage}[b]{0.5\linewidth}
\centering
\includegraphics[width=2.8in,height=2.3in]{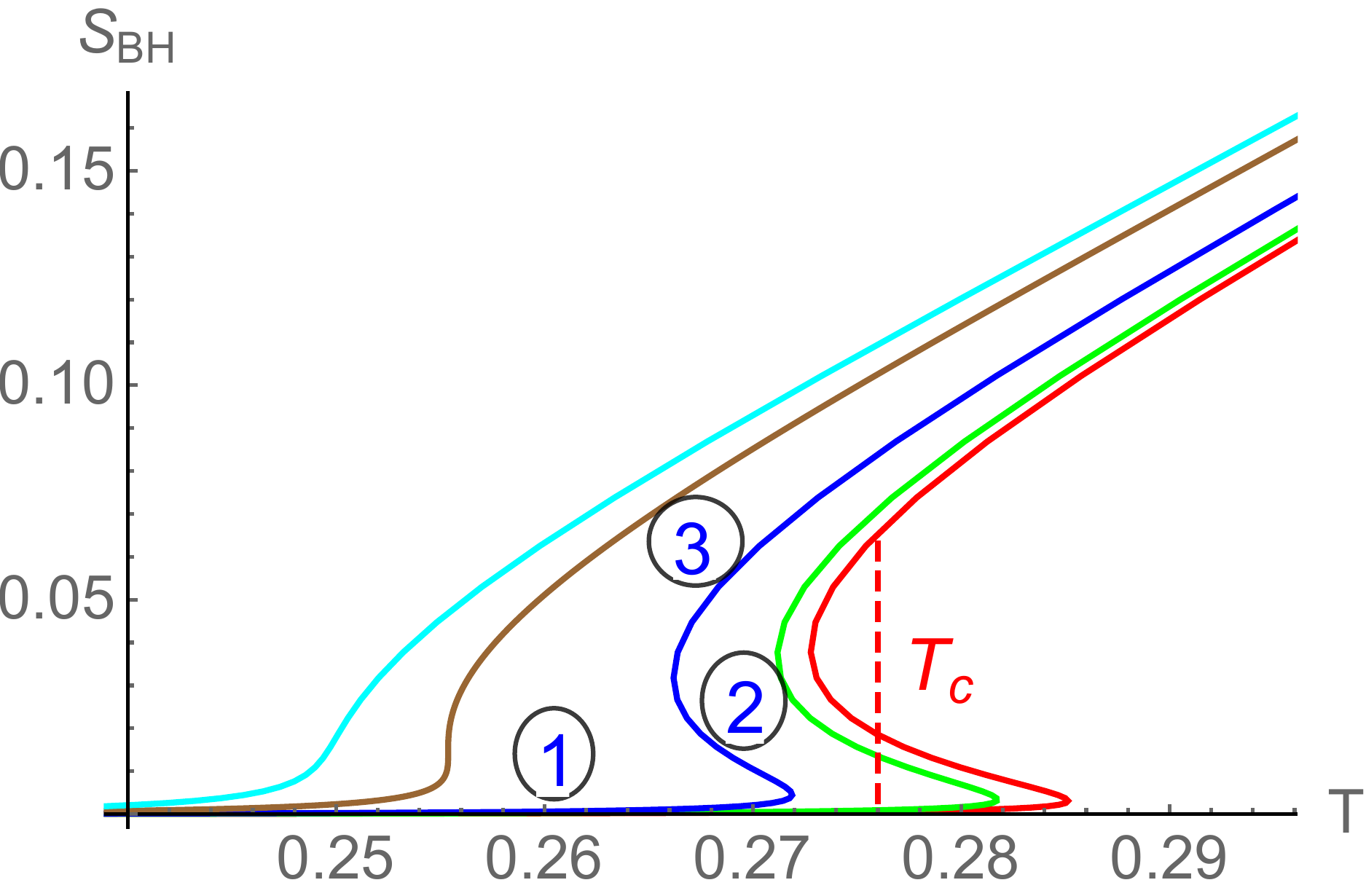}
\caption{ \small $S_{BH}$ as a function of $T$ for various values of the chemical potential $\mu$. Here red, green, blue, brown and cyan curves correspond to $\mu=0$, $0.1$, $0.2$, $0.312$ and $0.35$ respectively. In units \text{GeV}.}
\label{TvsSBHvsMucase2}
\end{minipage}
\hspace{0.4cm}
\begin{minipage}[b]{0.5\linewidth}
\centering
\includegraphics[width=2.8in,height=2.3in]{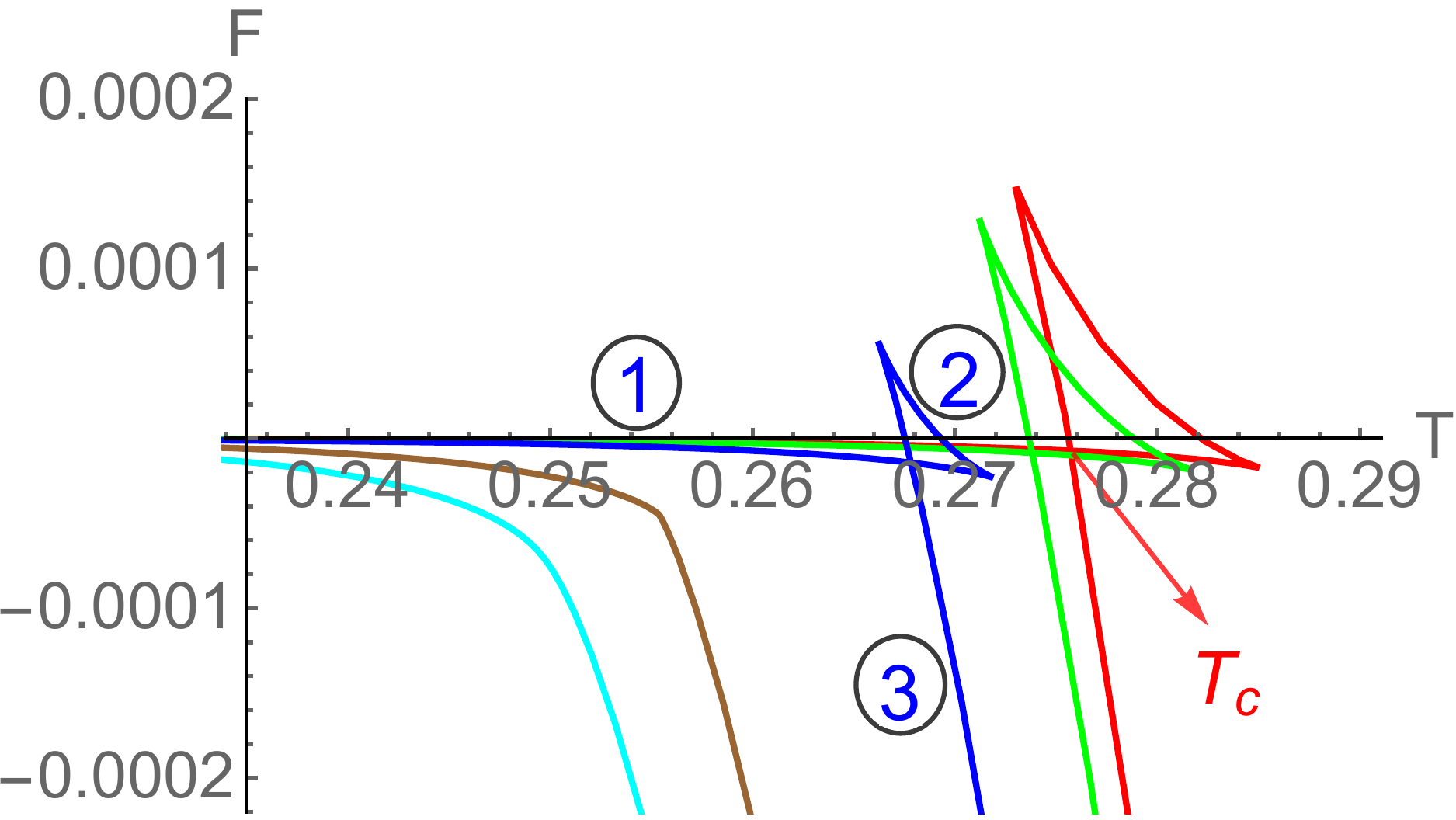}
\caption{\small $F$ as a function of $T$ for various values of the chemical potential $\mu$. Here red, green, blue, brown and cyan curves correspond to $\mu=0$, $0.1$, $0.2$, $0.312$ and $0.35$ respectively. In units \text{GeV}.}
\label{TvsFBHvsMucase2}
\end{minipage}
\end{figure}
The thermodynamics of the gravity solution (\ref{Aansatz2}) is shown in Figures~\ref{TvsSBHvsMucase2} and \ref{TvsFBHvsMucase2}. For small values of $\mu$, we now instead find three branches in the $(T,S_{BH})$ plane which are labeled by $\circled{1}$, $\circled{2}$ and $\circled{3}$. The branches $\circled{1}$ and $\circled{3}$ for which the entropy increases with temperature are stable whereas the branch $\circled{2}$ for which entropy decreases with temperature, is unstable. The stable branches $\circled{1}$ and $\circled{3}$ correspond to small (large $z_h$) and large (small $z_h$) size black holes respectively. This suggests a first order phase transition from a small to a large black hole phase as we increase the Hawking temperature. The phase transition and the corresponding critical temperature can be examined from the free energy behavior, which is shown in Figure~\ref{TvsFBHvsMucase2}. Here, the free energy is again normalised with respect to thermal-AdS. One can clearly observe a swallow-tail like structure in the free energy and a transition from small to large black hole phase at $T_{c}$. At $T_{c}$, the free energy of large black hole phase becomes smaller than the small black hole phase. For $\mu=0$, we find $T_{c}=0.276~\text{GeV}$. Interestingly, as opposed to the case with $A_{1}(z)$, now the black hole solution exists at all temperatures. In particular, the free energy of the stable black hole branches are always less than the thermal-AdS, indicating that black holes are more stable than thermal-AdS.\\

However, for higher values of $\mu$ the unstable branch starts decreasing in size and completely vanishes above a certain critical $\mu_c$. In particular, at $\mu_c=0.312~\text{GeV}$, small and large black holes join together to form a single black hole (shown by  brown line) which is stable and moreover exists at all temperatures. The critical temperature $T_{c}$ decreases with the chemical potential for $\mu\leq\mu_c$, at which the first order transition from small to large black hole phase ceases to exist. The full dependence of $T_{c}$ on $\mu$ and the complete phase diagram can be found in \cite{Dudal:2017max}. Overall we find a first order transition line, separating small and large black hole phases, terminating at the second order critical point $\mu_c$. This phenomenon is similar to the famous liquid-gas Van der Waals type phase transition. It is important to mention that in our model this Van der Waals type phase transition occurs with planar horizon as opposed to the spherical horizon case \cite{Hawking} which has been discussed to great extent in the literature \cite{Chamblin:1999tk,Chamblin:1999hg,Mahapatra:2016dae,Caldarelli:1999xj}. Similar kinds of a phase transition with planar horizon have been recently reported in \cite{Zhang:2017tbf,Anabalon:2016izw,Anabalon:2017yhv} in different contexts.

In \cite{Dudal:2017max}, we showed that this small/large black hole phase transition in the bulk corresponds to the specious-confined/deconfined phase transition in the boundary. In particular, by analysing the potential of the probe quark-antiquark pair, we showed that the large black hole phase is dual to deconfined phase whereas the small black hole phase is dual to specious-confined phase. Here,  specious-confined simply implies a phase which does not strictly correspond to the standard confined phase, however it shares many properties with the latter. For example, the specious-confined phase shows an area law for the Wilson loop at low temperatures and has an extremely small, however non-zero, Polyakov loop expectation value. Moreover, the free energy and entropy of the quark-antiquark pair in this specious-confined (as well as in the deconfined) phase are qualitatively similar to those of lattice QCD results. Even the temperature dependent behavior of the speed of sound in this specious-confined phase is in good agreement with lattice results.\\

Because of these fascinating properties of the dual boundary theory, which emerge from the scale factor $A(z)=A_{2}(z)$, it become important to investigate the entanglement entropy of this model as well. As we will show the entanglement entropy not only behaves differently in the specious-confined and deconfined phases but also highlights the delicate relation between specious-confined and standard confined phases.

\subsection{Holographic entanglement entropy}
Our aim in this subsection is to investigate the entanglement entropy in the above mentioned specious-confined/deconfined phases. For this purpose, we again consider a strip geometry of length $\ell$ as a subsystem. The necessary formulas are already derived in section $3$ and here we just present our numerical results.
\begin{figure}[h!]
\begin{minipage}[b]{0.5\linewidth}
\centering
\includegraphics[width=2.8in,height=2.3in]{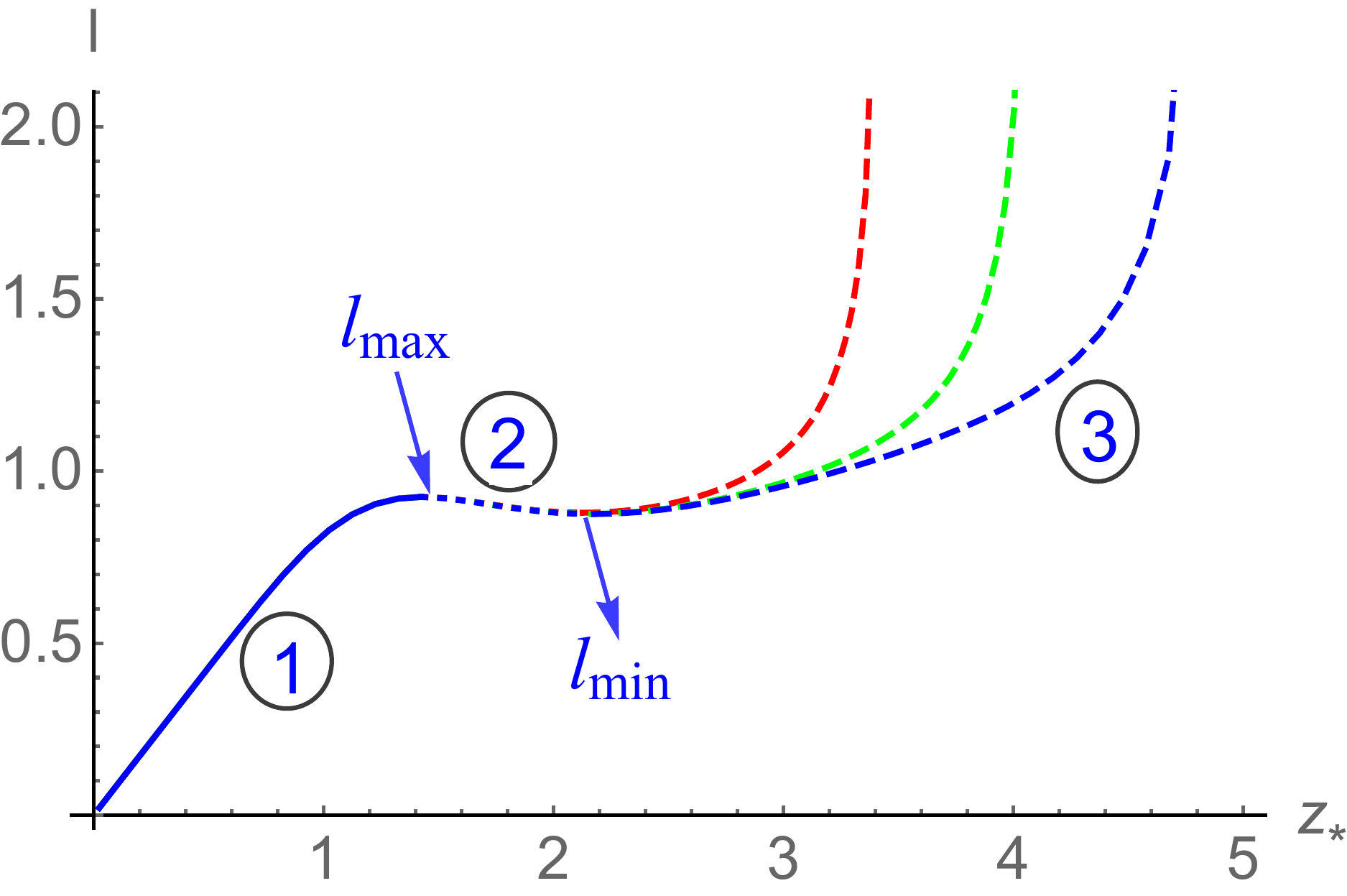}
\caption{ \small $\ell$ as a function of $z_*$ in the small AdS black hole background. Here $\mu=0$ and red, green and blue curves correspond to $T/T_{c}=0.9$, $0.8$ and $0.7$ respectively. In units \text{GeV}.}
\label{zsvslAdSBHMu0smallcase2}
\end{minipage}
\hspace{0.4cm}
\begin{minipage}[b]{0.5\linewidth}
\centering
\includegraphics[width=2.8in,height=2.3in]{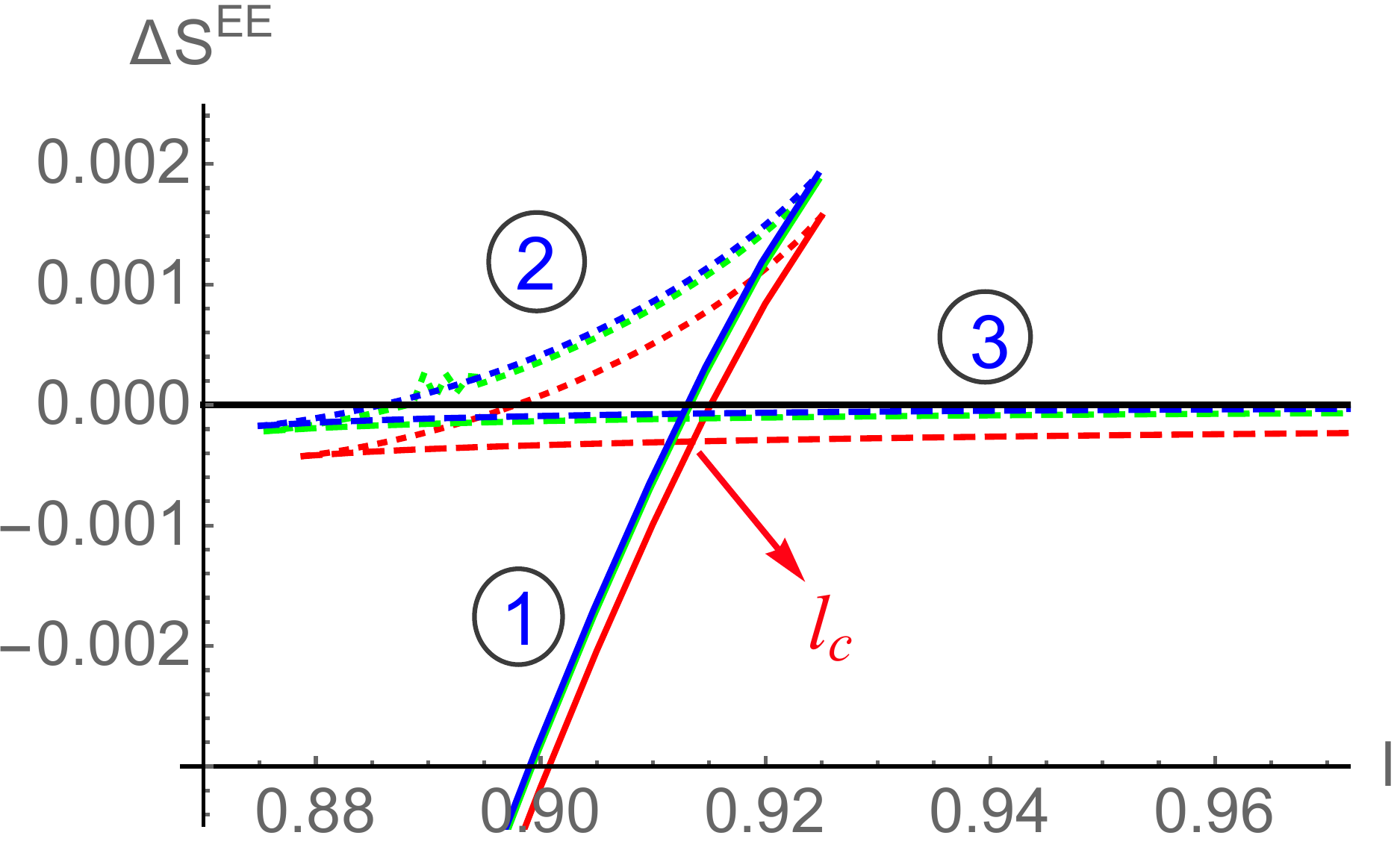}
\caption{\small $\Delta S^{EE}=S^{EE}_{con}-S^{EE}_{discon}$ as a function of $\ell$ in the small black hole background. Here $\mu=0$ and red, green and blue curves correspond to $T/T_{c}=0.9$, $0.8$ and $0.7$ respectively. In units \text{GeV}.}
\label{lvsSEEAdSBhMu0smallcase2}
\end{minipage}
\end{figure}
\\

Let us first discuss the entanglement entropy in the specious-confined phase whose dual gravity theory corresponds to a small black hole phase. The results for three different temperatures are shown in Figures~\ref{zsvslAdSBHMu0smallcase2} and \ref{lvsSEEAdSBhMu0smallcase2}. One can clearly observe a significant difference between specious-confined and standard confined phases. In particular, the $z_*$ vs.~$\ell$ behavior can now be divided into three branches instead of two. In the first branch, $\ell$ increases with $z_*$ till it reaches $\ell_{max}$; in the second branch it decreases from $\ell_{max}$ to $\ell_{min}$ and finally in the third branch it starts increasing again from $\ell_{min}$. These three branches are shown by solid, dotted and dashed lines and are marked by $\circled{1}$, $\circled{2}$ and $\circled{3}$ respectively. Interestingly, between $\ell_{min}$ and $\ell_{max}$, there are now three solutions for the connected surface for a given $\ell$. Therefore, although a certain $\ell_{max}$ does exist in the specious-confined phase, however as opposed to the standard confined phase, a connected solution now exists for all $\ell$ (remember that in the standard confined phase, as shown in the previous section, the connected surface exists only up to $\ell_{max}$). Consequently, the appearance of third branch (marked by $\circled{3}$) makes the transition between different entangling surfaces more non-trivial.\\

In Figure~\ref{lvsSEEAdSBhMu0smallcase2}, the difference between the entanglement entropy of the connected and disconnected surfaces is shown. Interestingly, we find a swallow-tail like structure in the entanglement entropy. The branch $\circled{2}$, for which $\ell$ decreases with $z_*$, makes the base of the swallow-tail and have a entanglement entropy which is always larger than the branches $\circled{1}$ and $\circled{3}$. This indicates that the branch $\circled{2}$ is actually a saddle point of the minimal area expression. To the best of our knowledge, such richness in the structure of entanglement entropy in holographic (QCD) models have not appeared in the literature before. Moreover, there is a phase transition from one connected surface (branch $\circled{1}$) to another connected surface (branch $\circled{3}$) as we increase $\ell$. This connected to connected surface transition is different from the connected to disconnected surface transition observed in the standard confined phase of the previous section. In particular, the number of degrees of freedom does not change due to this connected to connected surface transition,
\begin{eqnarray}
\frac{\partial S^{EE}}{\partial \ell} &\propto& \frac{1}{G_N} = \mathcal{O}(N^2)\quad\text{for both}\quad  \ell < \ell_{c}\quad \text{and}\quad   \ell > \ell_{c}\,.
\end{eqnarray}
We also note that the difference in the entanglement entropy of the connected and disconnected surfaces is always negative for any $\ell$. It suggests that even for very large subsystem size there is no transition from connected to disconnected surface. Consequently, the entanglement entropy is always of order $\mathcal{O}(N^2)$ for any subsystem size. From the entanglement entropy point of view, this is the biggest difference between specious-confined and standard confined phases. Whereas in the confined phase the order of the entanglement entropy changes at certain length $\ell_{c}$, no such thing happens in the specious-confined phase. This difference further underlines the subtle  interpretation of the small black hole phase as being strictly dual to the confined phase (a point overlooked in \cite{He:2013qq}, see also our discussion in \cite{Dudal:2017max}). Therefore, it seems that the entanglement entropy is not as efficient in probing the specious-confined phase as in probing the standard confined phase, we will come back to this issue momentarily . As we will show shortly, the entanglement entropy does however capture the essence of the small/large black hole phase transition of the gravity side.\\

To put matters into perspective, we like to emphasize here that, although $\frac{\partial S^{EE}}{\partial \ell}$ is non-zero in the specious-confined phase in a strict sense, it is very small. In particular, for larger subsystems, the entanglement entropy of the connected surface depends very mildly on $\ell$ and is in fact almost independent of it. This can be clearly seen from the third branch in Figure~\ref{lvsSEEAdSBhMu0smallcase2}. For example, we find that the change in the entanglement entropy happens only at the fifth decimal place when passing from $\ell$ to $\ell+\ell/2$. This behavior resembles closely ---however, not exactly--- the disconnected surface for which $\frac{\partial S^{EE}}{\partial \ell}=0$, highlighting again the subtle difference but also similarity between specious-confined and standard confined phases as was first mentioned in \cite{Dudal:2017max}.\\

For completeness,  we also like to mention that above results for the entanglement entropy are true irrespective of the value of the chemical potential. For even higher values of the chemical potential, we keep finding the above mentioned richness in the structure of entanglement entropy. In particular, the novel connected to connected surface transition is occurring. Similarly, the entanglement entropy for the connected surface is always smaller than the disconnected one and therefore no connected to disconnected surface transition is possible.
\\

To further quantify the specious-confined phase, it is instructive to consider the (UV finite) entropic $\mathcal{C}$-function \cite{Nishioka:2006gr}, as also studied numerically in \cite{Buividovich:2008kq,Itou:2015cyu}. This quantity, in our case defined as
\begin{equation}\label{Cf}
  \mathcal{C}(\ell)= \frac{\ell^3}{\text{Area}}\frac{\partial S^{EE}}{\partial \ell}\,,
\end{equation}
is sensitive to the number of degrees of freedom at length scale $\ell$, as such it is equally sensitive to the (de)confinement transition in terms of $\ell$ as described earlier in our paper, under the form of a sharp drop at $\ell_{c}$. This has been confirmed from lattice simulations \cite{Buividovich:2008kq,Itou:2015cyu} as well as from holographic viewpoint \cite{Dudal:2016joz}. The SU(3) lattice setup of \cite{Itou:2015cyu} suggested $\ell_{c}\approx 0.88~\text{fm}\approx 4.3~\text{GeV}^{-1}$, slightly larger than our first estimate of eq.~\eqref{lcr}. \\

\begin{figure}[h!]
\centering
\includegraphics[width=2.8in,height=2.3in]{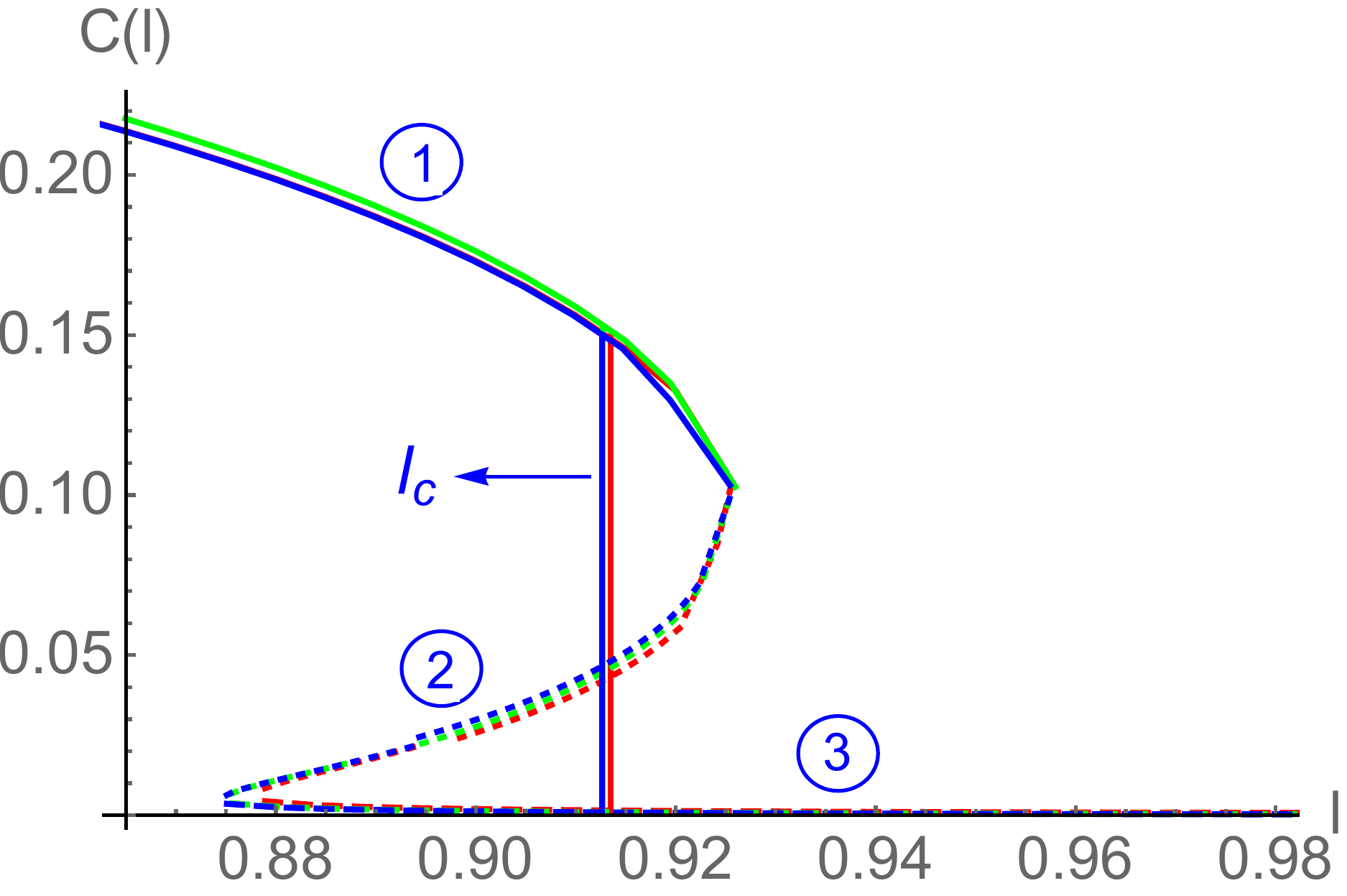}
\caption{ \small The entropic $\mathcal{C}$-function as a function of strip length $\ell$ in the specious-confined phase. Here $\mu=0$ and red, green and blue curves correspond to $T/T_{c}=0.9$, $0.8$ and $0.7$ respectively. In units \text{GeV}.}
\label{CvslsmallBHcase2}
\end{figure}

Returning to the specious-confined setup currently under scrutiny, we can equally well consider $\mathcal{C}(\ell)$, shown in Figure~\ref{CvslsmallBHcase2}, from which we can deduce an effective $\ell_{c}$ thanks to the sharp drop at the connected-connected surface transition (shown by vertical solid lines) and almost-zero value for larger values of $\ell$. At vanishing temperature and chemical potential, we find an estimate $\ell_{c}= 0.913 \ \text{GeV}^{-1}$, rather close to our earlier value \eqref{lcr} making use of the other, simpler form factor $A_1(z)$.  Moreover, the expected behavior that the magnitude of $\mathcal{C(\ell)}$ should decreases monotonically as the size of the entangling surface increases, i.e.~from UV to IR, is also evident form Figure~\ref{CvslsmallBHcase2}. This behavior can be seen for both stable branches ($\circled{1}$ and $\circled{3}$). Since $\mathcal{C(\ell)}$ measures the degrees of freedom in a system at the energy scale $\frac{1}{\ell}$, we see that $\mathcal{C(\ell)}$ decreases under the RG-flow in our specious-confined setup as well. \\

\begin{figure}[h!]
\begin{minipage}[b]{0.5\linewidth}
\centering
\includegraphics[width=2.8in,height=2.3in]{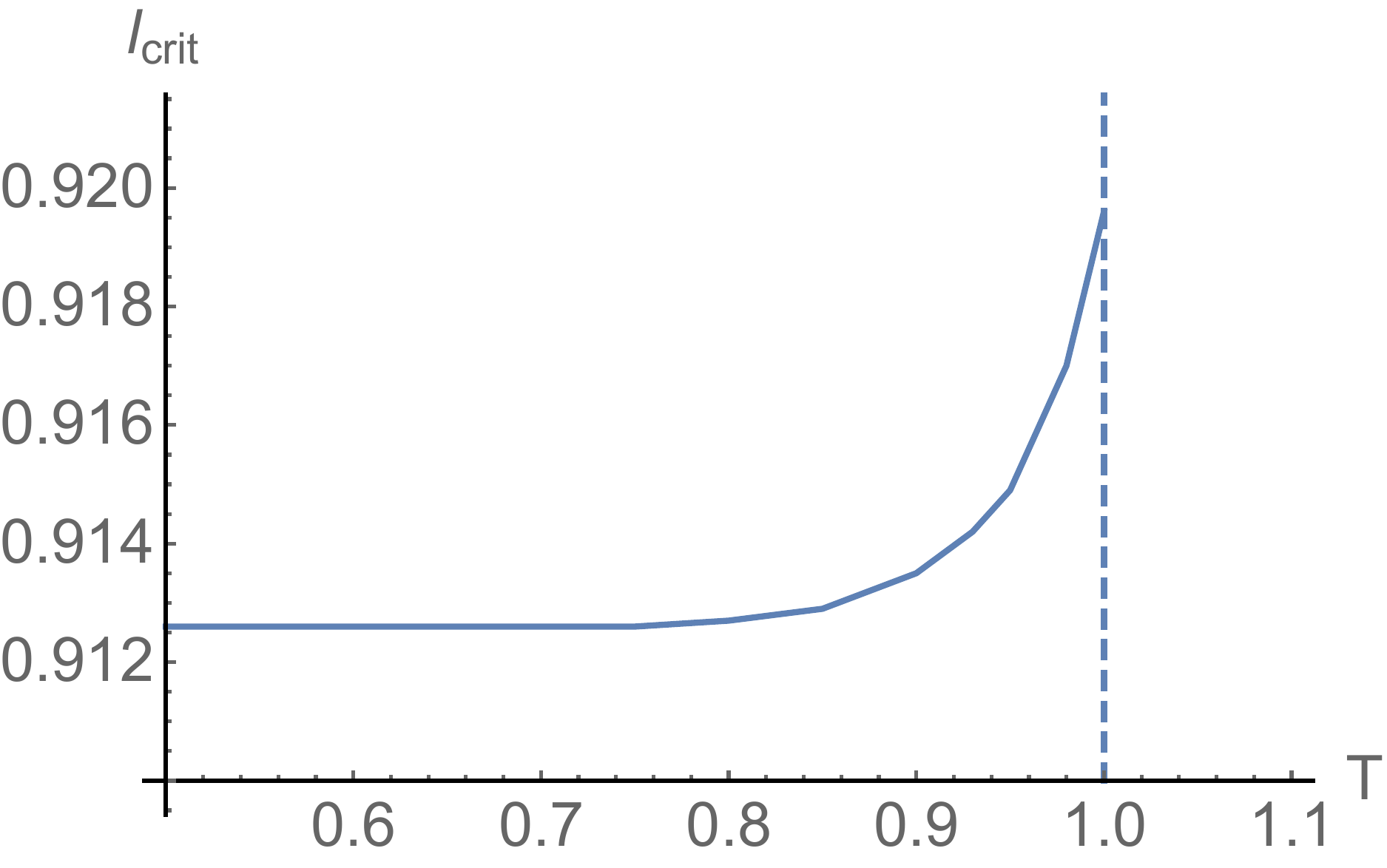}
\caption{ \small $\ell_{c}$ as a function of $T$ in the in the specious-confined phase at $\mu=0$. This $(T, \ell_{c})$ holographic phase diagram can be compared with the SU(2) lattice gauge theory conjecture of \cite[Figure 8]{Buividovich:2008kq}. In units \text{GeV}.}
\label{lcritvsTMu0smallBHcase1}
\end{minipage}
\hspace{0.4cm}
\begin{minipage}[b]{0.5\linewidth}
\centering
\includegraphics[width=2.8in,height=2.3in]{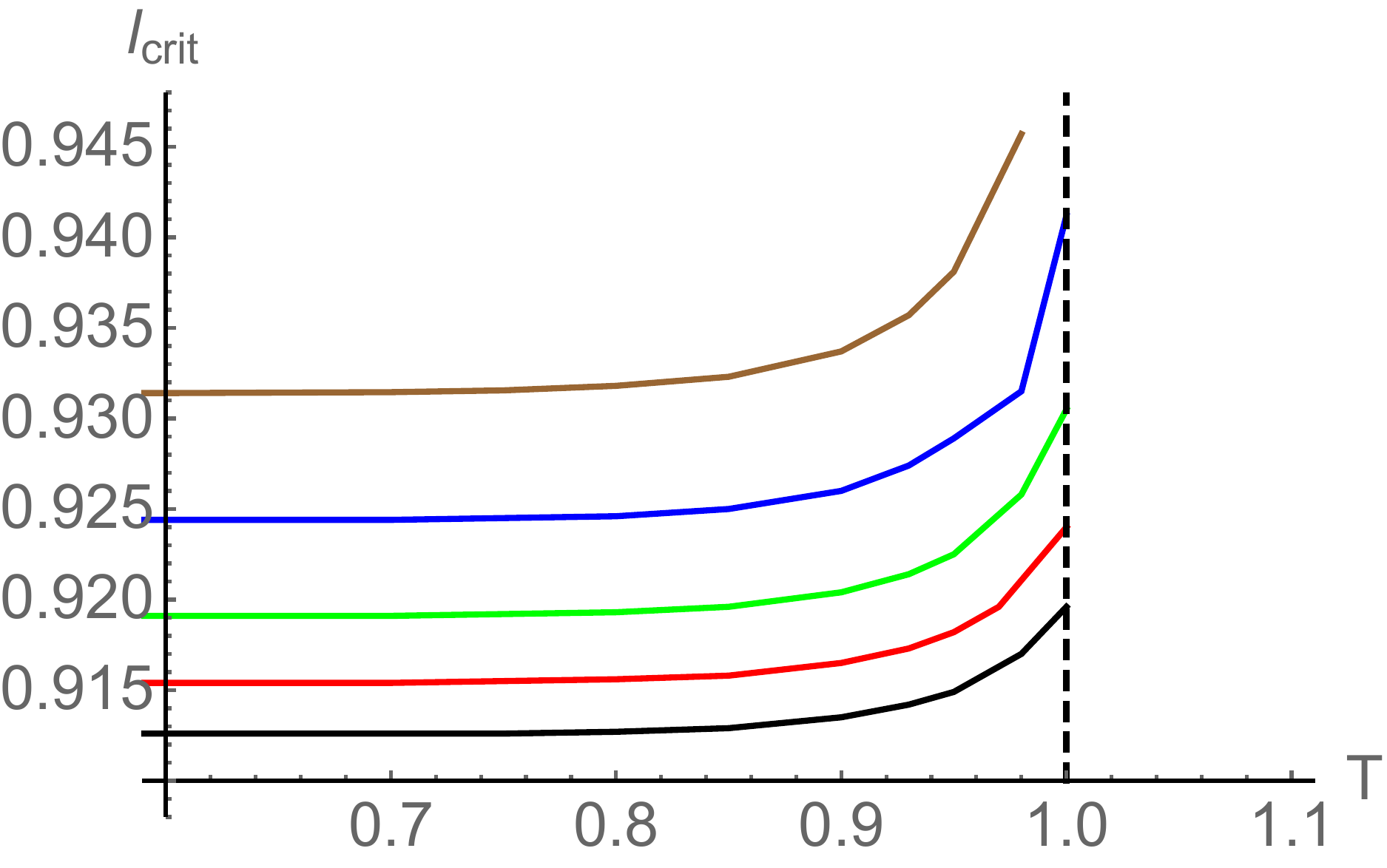}
\caption{\small $\ell_{c}$ as a function of $T$ in the specious-confined phase. Here red, green, blue and brown curves correspond to $\mu=0.10$, $0.15$, $0.20$ and $0.25$ respectively. The black curve is for $\mu=0$, and is shown for comparison. In units \text{GeV}.}
\label{lcritvsTvsMusmallBHcase1}
\end{minipage}
\end{figure}

More interestingly, in the specious-confined setup, we can switch on temperature and chemical potential in the confined phase and probe the response of the entanglement entropy, or more precisely of the $\mathcal{C}$-function to it when the entangling surface grows. This allows to draw a $(T,\mu,\ell)$ phase diagram, see Figures~\ref{lcritvsTMu0smallBHcase1} and \ref{lcritvsTvsMusmallBHcase1}, which to the best of our knowledge has not appeared in literature so far, apart from the conjectured phase diagram in \cite[Figure 8]{Buividovich:2008kq}. For clarity, we have also drawn the $(T,\ell)$ phase diagram for various values of $\mu$ in Figure~\ref{lcritvsTvsMusmallBHcase1}. We note that for fixed $\mu$, $\ell_{c}$ first increases slowly with $T$, while experiencing a sharp rise until its ending at $T=T_c$. On general grounds, we do expect a rising $\ell_c$, as the more the system gets heated up towards deconfinement, the lesser confining to becomes, so the larger the subsystem will need to grow to again experience confinement in the entanglement entropy due to sampling over the outer degrees of freedom.
\\
\\
In Figure~\ref{lcritvsTvsMusmallBHcase1}, we have shown the variation $\ell_c$ with $T$ in the specious confined phase for various values of $\mu$. We find that the magnitude of $\ell_{c}$ decreases with $T$ and approaches a $\mu$ dependent constant value at vanishing temperature. We do not have lattice QCD results here to compare with, and in this regard, our result in Figure~\ref{lcritvsTvsMusmallBHcase1} can be considered as a genuine prediction from holographic QCD. \\

\begin{figure}[h!]
\begin{minipage}[b]{0.5\linewidth}
\centering
\includegraphics[width=2.8in,height=2.3in]{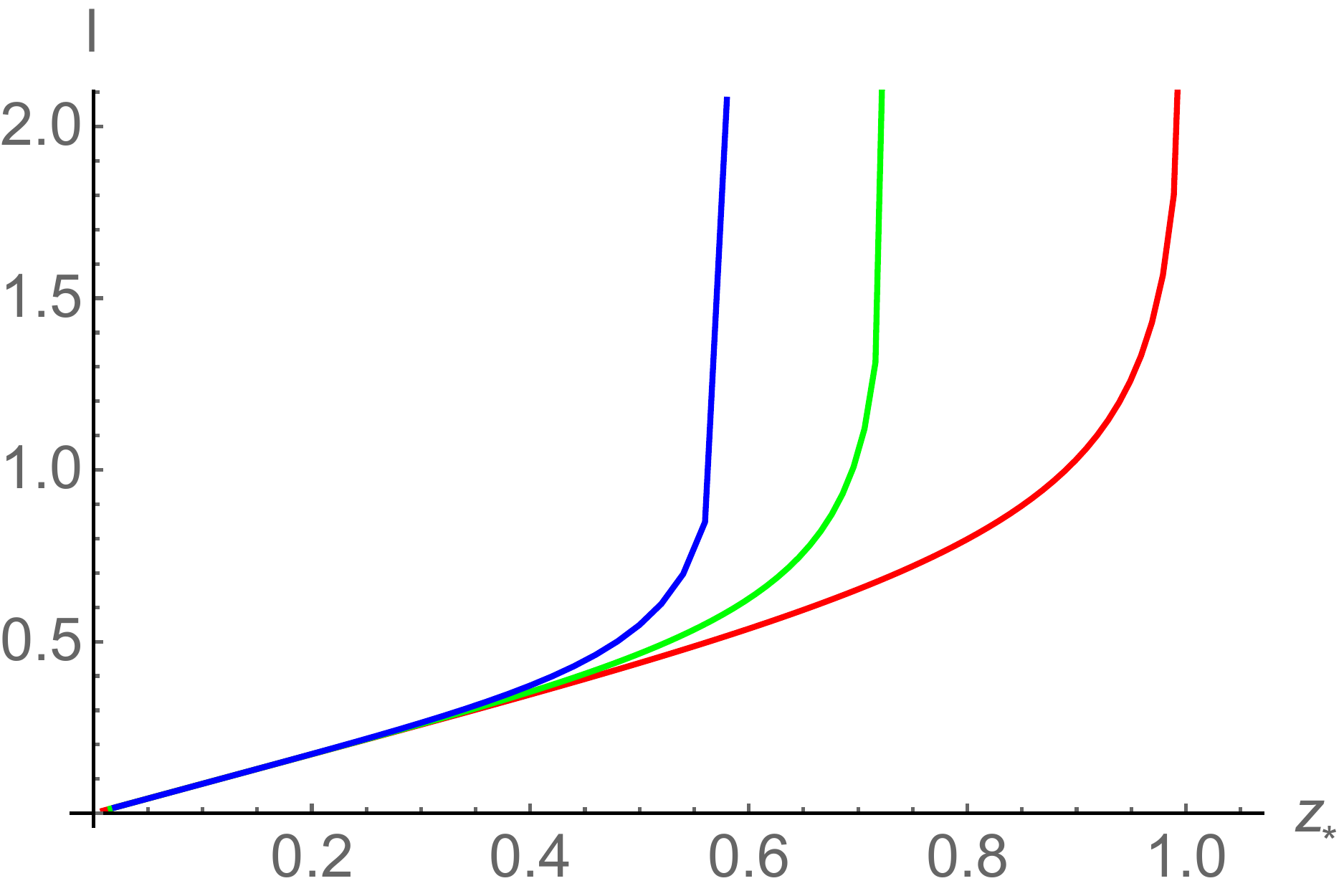}
\caption{ \small $\ell$ as a function of $z_*$ in the large AdS black hole background. Here $\mu=0$ and red, green and blue curves correspond to $T/T_{c}=1.2$, $1.6$ and $2.0$ respectively. In units \text{GeV}.}
\label{zsvslAdSBHMu0largecase2}
\end{minipage}
\hspace{0.4cm}
\begin{minipage}[b]{0.5\linewidth}
\centering
\includegraphics[width=2.8in,height=2.3in]{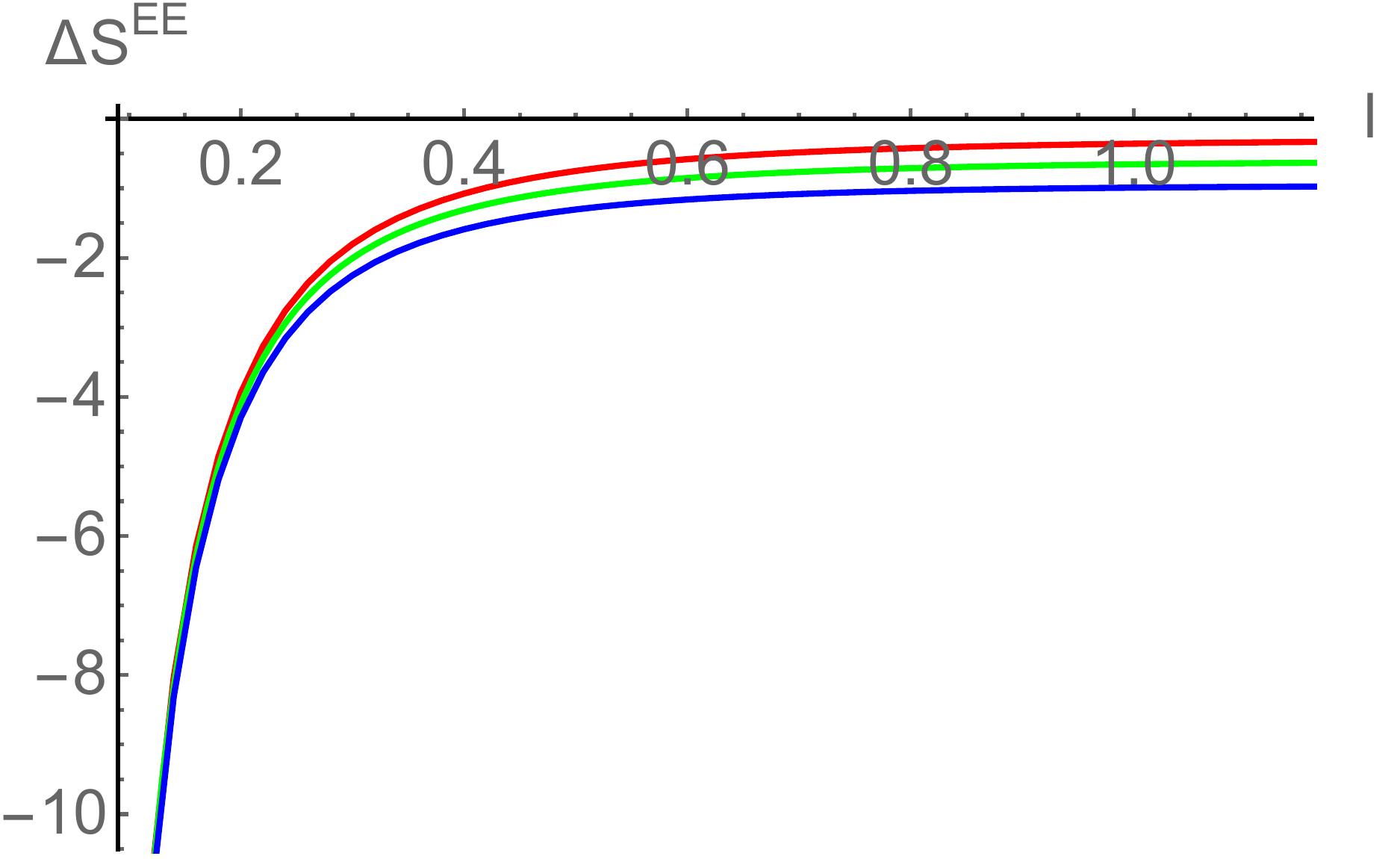}
\caption{\small $\Delta S^{EE}=S^{EE}_{con}-S^{EE}_{discon}$ as a function of $\ell$ in the large black hole background. Here $\mu=0$ and red, green and blue curves correspond to $T/T_{c}=1.2$, $1.6$ and $2.0$ respectively. In units \text{GeV}.}
\label{lvsSEEAdSBhMu0largecase2}
\end{minipage}
\end{figure}

\begin{figure}[h!]
\centering
\includegraphics[width=2.8in,height=2.3in]{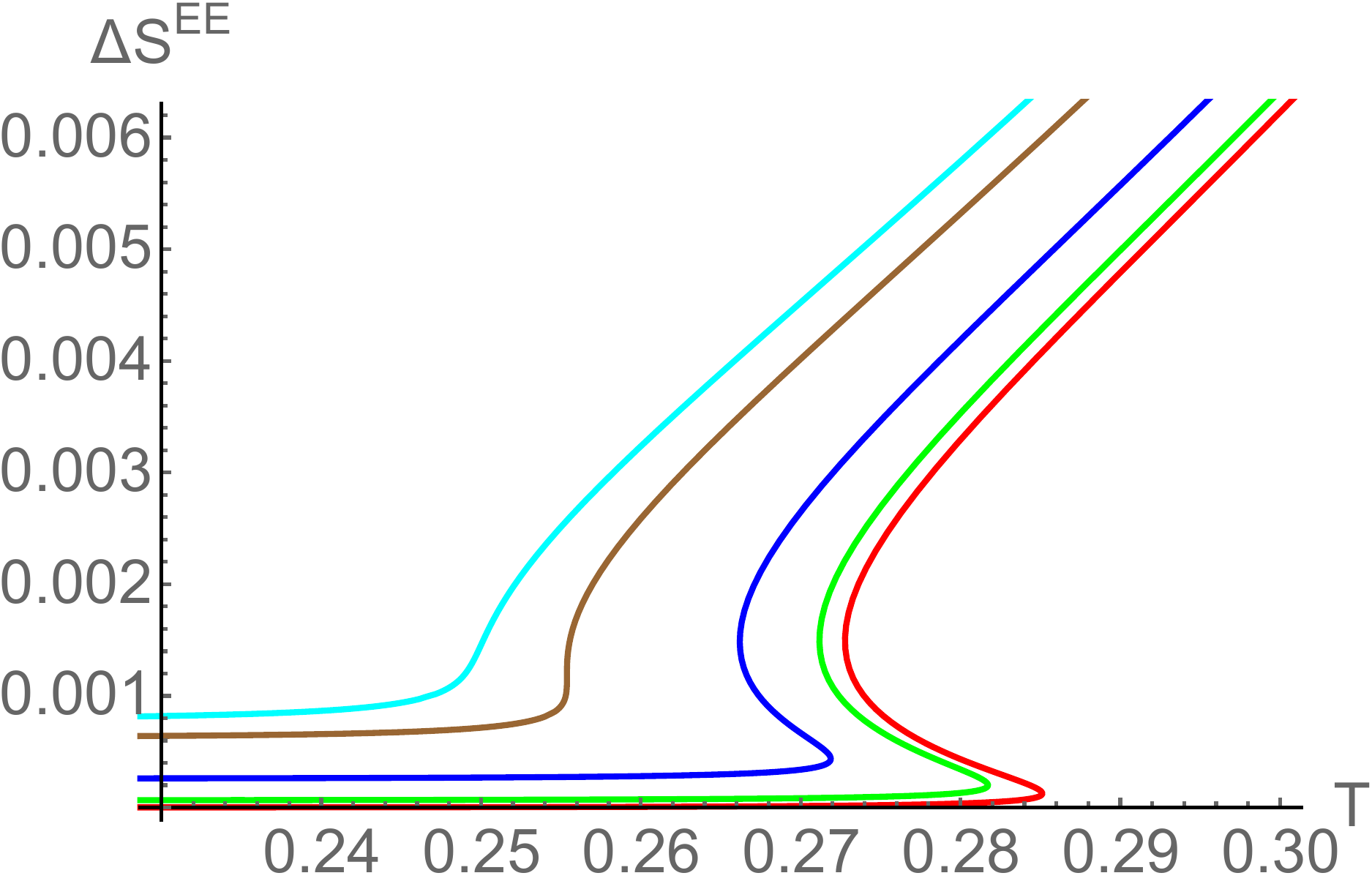}
\caption{ \small $\Delta S^{EE}=S^{EE}_{AdS-BH}-S^{EE}_{Thermal \ AdS}$ as a function of $T$. Here $\ell=0.2$ and red, green, blue, brown and cyan curves correspond to $\mu=0.0$, $0.1$, $0.2$, $0.312$ and $0.35$ respectively. In units \text{GeV}.}
\label{TvsdelSEEvsMucase2}
\end{figure}

Next, we discuss the entanglement entropy in the deconfined phase whose dual gravity theory corresponds to a large black hole phase. The results, shown in Figures~\ref{zsvslAdSBHMu0largecase2} and \ref{lvsSEEAdSBhMu0largecase2}, are quite similar to those of deconfined phase results of the previous section (where $A(z)=A_{1}(z)$), and therefore we can be brief here. Again, the connected entangling surface exists for all $\ell$ and there is a one to one relation between $z_*$ and $\ell$, suggesting that for larger and larger subsystem size the turning point of the connected surface moves closer towards the horizon. This means that for larger subsystem size the connected entangling surface looks more like a disconnected one. Moreover, neither $\ell_{max}$ nor $\ell_{c}$ exist and hence no phase transition between different entangling surfaces. Consequently, as in the previous section, the entanglement entropy is always of order $\mathcal{O}(N^2)$ in the deconfined phase.\\
\\
Finally, we discuss the thermodynamical aspects of the entanglement entropy and its connection with the thermal entropy of the small and large black holes. Our main aim here is to investigate whether the entanglement entropy can also capture the difference between the small/large black hole (thus between the dual specious-confined/deconfined) phase transition as for the thermal-AdS/black hole (or the dual confined/deconfined) phase transition in the previous section. The results are shown in Figure~\ref{TvsdelSEEvsMucase2} for a fixed $\ell=0.2~\text{GeV}^{-1}$, although the essential features of our analysis remain unchanged even for other values of $\ell$. We again find a striking similarity in the structure of entanglement entropy and thermal entropy of black holes (shown in Figure~\ref{TvsSBHvsMucase2}). In particular, there are again three branches in the entanglement entropy for small values of $\mu$. The branches with positive slope correspond to the stable solutions whereas the branch with negative slope corresponds to the unstable one. The two stable branches coincide to small and large black hole phases. Moreover, the unstable solution does not exist above $\mu_c=0.312~\text{GeV}$ and there is only branch which is stable at all temperatures (shown by the cyan line). As such, the entangling surface, which propagates in the bulk, again experiences the effects caused by the changing geometry of the black hole phase transitions.  These similarities again highlight the effectiveness of entanglement entropy to probe black hole and dual specious-confined/deconfined phase transition, however yet again, without providing any information about the critical temperature.

\section{Conclusions}
We have further investigated how the entanglement entropy can shed new light on the QCD confinement mechanism. In order to do so, we employed the holographic entanglement entropy dictionary of Ryu--Takayanagi, applied to a recently introduced dual QCD model \cite{Dudal:2017max} which has the important property of being a consistent gravitational model that includes a stable small and large black hole phase, and depending on the temperature, one is thermodynamically dominant over the other. We have essentially recovered these features in the corresponding entanglement entropies, constituting a nice result on the gravitational side. Noteworthy is a novel type of transition between entangling surfaces in the small black hole phase, namely between 2 different connected surfaces. This appears to be the first time such transition is reported about in the literature. It corresponds to a cusp in the entanglement entropy, albeit that the order of degree of freedom counting does not change as it concerns the same type of surface (both closed). \\

On the dual QCD side, these two different black hole phases correspond to either a confinement-like phase (small black hole) or deconfinement phase (large black hole). In \cite{Dudal:2017max}, we already scrutinized this confinement-like phase in terms of the Wilson and Polyakov loops, next to the interquark free energy and thermal entropy. Indeed, the phase being dual to a black hole immediately allows to introduce temperature even in the confinement phase, adding an extra ingredient when compared to other holographic models already on the market. Depending on which QCD properties one wishes to describe exactly, one or another holographic model becomes more appropriate. To our knowledge, there is no dual QCD model yet capable of describing all of QCD's (non-)perturbative physics over all possible temperature ranges, evidently also not ours. \\

In the current paper, we have given further credit to our model by investigating its entanglement entropy structure, confirming its interpretation as being a decent dual description of QCD above and below deconfinement. This becomes best visible from the behavior of the entropic $\mathcal{C}$-function. Perhaps the most striking result to report about is the first estimate for the QCD phase diagram in terms of temperature, chemical potential and the entangling surface's strip length, summarized in our Figures~\ref{lcritvsTMu0smallBHcase1} and \ref{lcritvsTvsMusmallBHcase1}.\\

The next step in our research setup will be to add a background magnetic field to it, while maintaining the validity of the Einstein gravitational equations of motion. If we succeed in doing so, we can tackle several important research questions about magnetized QCD using holography once more, where the magnetic field will leave its anisotropic footprints in many relevant quantities, including the entanglement entropy \cite{Dudal:2016joz}. In addition, even without magnetic field, having access to a temperature-dependent confinement phase, our current model can be further used to study QCD around the phase transition, including the behavior of heavy meson spectral functions, let us only mention the melting properties of quarkonia. \\

At last, we hope that these seminal predictions of our model, in particular of the phase diagram, can also inspire further lattice QCD investigations of entanglement entropy, perhaps following the lines of \cite{Caselle:2018kap}.

\section*{Acknowledgments}
The work of S.~M.~was partially supported by a PDM grant of KU Leuven. The work of S.~M.~is supported by the Department of Science
and Technology, Government of India under the Grant Agreement number IFA17-PH207 (INSPIRE Faculty Award).

\end{document}